\documentclass{article}
\usepackage[utf8]{inputenc}
\usepackage[T1]{fontenc}
\usepackage{geometry}
\usepackage{float}
\usepackage{color}
\usepackage{array}
\usepackage{multirow}
\usepackage{amstext}
\usepackage{graphicx}
\usepackage{nicefrac}
\usepackage{inputenc}
\usepackage[spanish]{babel}
\usepackage{authblk}
\usepackage{ulem}
\usepackage{nomencl}

\makenomenclature

\usepackage{etoolbox}
\renewcommand\nomgroup[1]{%
  \item[\bfseries
  \ifstrequal{#1}{A}{Parameter name}{%
  \ifstrequal{#1}{B}{Abbreviations}{%
  }}%
]}

\date{}

\begin{document}

\title{Energetic optimization considering a generalization of the ecological criterion in traditional simple-cycle and combined cycle power plants}

\author[$1$]{S. Levario--Medina}
\author[$1$]{G. Valencia--Ortega}
\author[$2,*$]{M. A. Barranco--Jiménez}
\affil[$1$]{Departamento de F\'{i}sica, Escuela Superior de F\'{i}sica y Matem\'{a}ticas, Instituto Polit\'{e}cnico Nacional, U. P. Zacatenco, Edif. 9, 2o Piso, Ciudad de M\'{e}xico, 07738, M\'{e}xico.}
\affil[$2$]{Escuela Superior de C\'{o}mputo del Instituto Polit\'{e}cnico Nacional, Av. Miguel Bernard, Esq. Av. Miguel Oth\'{o}n de Mendizabal, Colonia Lindavista, Ciudad de M\'{e}xico 07738, M\'{e}xico.

$^1$levario@esfm.ipn.mx
$^1$gvalencia@esfm.ipn.mx
$^*$mbarrancoj@ipn.mx}

\maketitle

The fundamental issue in the energetic performance of power plants, working both as traditional fuel engines and as combined cycle turbine (gas-steam), lies in quantifying the internal irreversibilities which are associated with the working substance operating in cycles. The purpose of several irreversible energy converter models is to find objective thermodynamic functions that determine operation modes for real thermal engines and at the same time study the trade off between energy losses per cycle and the useful energy. As those objective functions, we focus our attention on a generalization of the so-called ecological function in terms of an $\epsilon$--parameter that depends on the particular heat transfer law used in the irreversible heat engine model. In this work, we mathematically describe the configuration space of an irreversible Curzon-Ahlborn type model. The above allows to determine the  optimal relations between the model parameters so that a power plant operates in physically accessible regions, taking into account internal irreversibilities, introduced in two different ways (additively and multiplicatively). In addition, we establish the conditions that the $\epsilon$--parameter must fulfill for the energy converter works in an optimal region between maximum power output and maximum efficiency points.

\mbox{}

\nomenclature[A, 01]{$T_h$}{Hot reservoir temperature ($K$)}
\nomenclature[A, 02]{$T_c$}{Cold reservoir temperature ($K$)}
\nomenclature[A, 04]{$T_{hw}$}{Hot internal reservoir temperature ($K$)}
\nomenclature[A, 05]{$T_{cw}$}{Cold internal reservoir temperature ($K$)}
\nomenclature[A, 08]{$\alpha$}{Thermal conductance $\left(\nicefrac{MW}{K}\right)$}
\nomenclature[A, 09]{$\beta$}{Thermal conductance $\left(\nicefrac{MW}{K}\right)$}
\nomenclature[A, 10]{$\gamma$}{Thermal conductances ratio(-)}
\nomenclature[A, 11]{$\delta$}{Heat leak's thermal conductance $\left(\nicefrac{MW}{K}\right)$}
\nomenclature[A, 03]{$\tau$}{Reservoirs temperature ratio (-)}
\nomenclature[A, 12]{$\sigma_i$}{Internal entropy production ($MW$)}
\nomenclature[A, 13]{$r$}{Irreversible parameter for CA-r model (-)}
\nomenclature[A, 14]{$R$}{Irreversible parameter for CA-R model (-)}
\nomenclature[A, 06]{$a_h$}{High reduced temperature (-)}
\nomenclature[A, 07]{$a_c$}{Low reduced temperature (-)}
\nomenclature[A, 15]{$\epsilon$}{Parameter of the generalization of the ecological function}
\nomenclature[B, 01]{$E$}{Ecological function}
\nomenclature[B, 02]{$E_G$}{Generalization of the ecological function}
\nomenclature[B, 03]{$MP$}{Maximum power output point}
\nomenclature[B, 04]{$M\eta$}{Maximum efficiency point}
\nomenclature[B, 05]{$ME_G$}{Maximum generalized ecological function point}
\nomenclature[B, 06]{$Z_i$}{Operation zones ($i=I,II,III$)}
\nomenclature[B, 07]{$HP$}{High power output}
\nomenclature[B, 08]{$LP$}{Low power output}
\nomenclature[B, 09]{$HE$}{High efficiency}
\nomenclature[B, 10]{$LE$}{Low efficiency}
\nomenclature[B, 11]{$HD$}{High dissipation}
\nomenclature[B, 12]{$LD$}{Low dissipation}

\printnomenclature

\section{Introduction}
In recent decades, various areas of the knowledge related to production of non fossil fuels and sustainable generation of energy have invested efforts in combining innovative operation cycles with waste recovery heat systems \cite{Chen99,Chen10,Feidt18}. As a result, numerous studies have been developed to determine energy conversion processes that reflect the best trade off between the maximum useful power generated and the maximum achievable efficiency \cite{Velascoetal,SanchezOrgaz13,MJSantos16,Levario19}. In general, the study of thermal engines has allowed not only to design more sophisticated engines but also to focus the attention on a more flexible operation, since the emissions during the combustion have become increasingly low. The type of thermal engines (energy converters), known as power plants have diversified and evolved due to social needs, whether for ecological \cite{Velascoetal} or economic \cite{BejanLibro,AngBarInst07} reasons. In this context, the combined cycle power plants in whose construction coexist two thermodynamic cycles from the same source of heat, have turned out to have a greater amount of available energy. Although in practice, the amount of non useful generated energy compared to the used one continues to be a big problem, it can be modulated (reduced) by paying attention to the operation regimes \cite{SilvArias13,Barranco14,GonzAyala17} with which this type of power plants can be operated. These performance regimes are normally associated with operation and design  parameters that measure roughly the internal and external irreversibilities.

There are several branches of non equilibrium thermodynamics \cite{Feng11,Feidt17} in which physical models have been established to understand the performance of energy converters. In particular, we have used the approximation of Finite Time Thermodynamics (FTT) \cite{Hoffmann97,Wu99,Durmayaz04}. Although, this approximation does not consider all of the features of certain thermodynamic systems as in other branches of non equilibrium thermodynamics \cite{Petrescu02,Sun11,Feidt13}, it has shown that FTT models have reproduced in a good way, several observed results related to dynamic models \cite{Levario19,AnguloJ94,AnguloR96,HoffmannOtto,Curtoetal,CurtoM09,Izumida08,Rojas18}, which include the energy losses due to heat conduction and frictional losses. Therefore, FTT models are good option to emulate energy conversion processes by means of objective thermodynamic functions in several irreversible power plant models \cite{HoffmannOtto,Curtoetal,DeVosLibro,ParaR2,ParaR1}. Within this context, different parameters related to construction and operation of the energy converters play a fundamental role to characterize objective functions which achieve a good trade off for process variables such as power output, efficiency and dissipation  \cite{Angulo91,Calvo01,Yilmaz}. One of the most used model within the FTT context is the endoreversible Curzon and Ahlborn model (CA model) \cite{CAMod}. This model allows us to establish upper limits for the operation modes that a thermal engine can undergoes considering not only external irreversibilities (heat transfer laws) but also the ones inside the working substance \cite{Andresen2001}.

In 1994, as an extension to the CA model \cite{Rubin} was proposed by Özkaynak et al \cite{ParaR2} and Chen \cite{ParaR1}, where the non-endoreversibility lumped parameter $R$ was introduced. It takes into account the irreversibility degree of internal processes within the working substance. Another way of quantifying internal irreversibilities is through the so-called uncompensated Clausius heat $r$, firstly proposed by Tolman and Fine \cite{Tolman}, later used by Silva-Martinez, Arias-Hernandez \cite{SilvArias13} and recently by Levario-Medina \cite{LevarioArias19}. This $r$ parameter roughly measures the non recovered amount of heat during the operation cycles. This way of introducing dissipative effects was firstly proposed by Bejan, Gordon and Huleihil \cite{Bejan,Gordon1,Gordon3}, they verified that adding the information of a heat bypass, a more real behavior is reproduced in the operation of thermal engines (the well known loops in the power output versus efficiency space).

In a recent paper \cite{Levario19}, the energetics of the CA model was studied by means of a generalization of an objective function called efficient power \cite{Yilmaz,JEI2009}, with the aim of obtaining physically accessible operation points for thermal engines. With the help of the extremal properties, which are the generalization of the known objective functions within the context of FTT, the best performance conditions can be found in terms of the design and construction parameters of each heat engine. The obtained conditions for each energy converter (power plants) delimit  an energetic zone with high power output and high efficiency. In case of the ecological function \cite{Angulo91}, it was shown that its optimization leads us to get a better economic performance than the efficient power one \cite{JEI2009}, since it has the implicit idea of obtaining the highest possible power output and the highest efficiency at the lowest energetic costs (at lower entropy production).

In this work, we consider a CA heat engine model with a heat leak and considering two different internal irreversibilities (the CA--like case with $R$-internal irreversibility and the CA--like case with $r$-internal irreversibility). By means of a generalization of the ecological function \cite{GFE1}, we identified three well-defined operation zones ($ZI$, $ZII$ and $ZIII$) in the obtained characteristic loops for irreversible heat engine model. Those zones are completely characterized in the power output-efficiency configuration space by the generalization parameter ($\epsilon$). The paper is organized as follows: in Section 2, we mathematically described the irreversible CA model's cases, as well as the conditions to reach the optimum points for both regimes: the maximum power output and maximum efficiency ones. In Section 3, we studied the constraints that must be satisfied the parameters associated with the irreversibilities. In the irreversible model, we established the operation zones ($ZI$, $ZII$ and $ZIII$) of the power plants. We also showed that the $\epsilon$-parameter can modify the operation conditions for every power plant. Finally, in Section 4, we present our conclusions.

\section{Energetic description of a CA--like heat engine}
\begin{figure}
	\centering
		\includegraphics[scale=1]{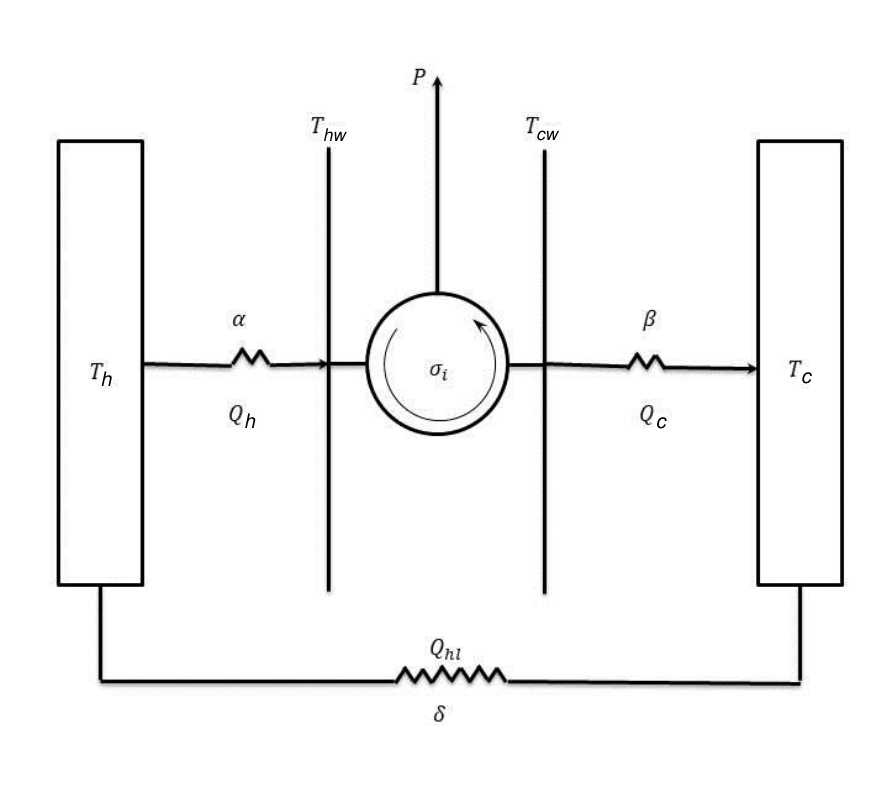}
	  \caption{Sketch of a modified CA heat engine}
	\label{fig:MoTiCuAl}
\end{figure}

In this section, we present a thermodynamic analysis of a Curzon-Ahlborn-like irreversible engine extended as shown in Fig. \ref{fig:MoTiCuAl}, which consists of two energy reservoirs; the first one at temperature $T_{h}$ and the other one at temperature $T_{c}$, where $T_{h}>T_{c}$. As is well known, the CA model includes two auxiliary energy reservoirs with working temperatures $T_{hw}$ and $T_{cw}$, with $T_{h}>T_{hw}>T_{cw}>T_{c}$ and they are in contact with a working substance operating in cycles. As in the typical CA model for a heat engine, this variant incorporates two thermal conductances $\alpha>0$ and $\beta>0$ that reflect the existence of natural heat flows through the materials that make up the heat exchangers. Besides, in this version of the CA--heat engine model, another thermal conductance ($\delta>0$) for the heat leak and in addition the internal irreversibilities within the working substance are taken into account by means of an appropriate parameter, this model emulates a behavior closer to what is happening in real heat engines.

From Fig. \ref{fig:MoTiCuAl}, if we consider a linear heat transfer law (Newtonian law) the heat fluxes are given by:

\begin{equation}
Q_{h}=\alpha T_{h}\left(1-a_{h}\right),\label{eq:Qh}
\end{equation}
\begin{equation}
Q_{c}=T_{h}\frac{\alpha\tau}{\gamma}\left(\frac{1}{a_{c}}-1\right),\label{eq:Qc}
\end{equation}
and
\begin{equation}
Q_{hl}=T_{h}\delta\left(1-\tau\right),\label{eq:Qhl}
\end{equation}
where $\gamma=\alpha/\beta$, $\tau=T_{c}/T_{h}$ ($0<\tau<1$); $a_{h}=T_{hw}/T_{h}$ is defined as the high reduced temperature and  $a_{c}=T_{c}/T_{cw}$ as the low reduced temperature.

It is well known that the performance in several types of energy converters has become a topic of general interest, and its study has been carried out within different contexts \cite{Feidt17,SalamonLibro,GabrielArias}. FTT has allowed to establish mathematical relationships that contain information on the way in which energy exchanges take place between the system and its surroundings through phenomenological parameters. These relationships have derived in the main process variables such as power output, efficiency and dissipation. In this model, these process functions have the following form:

\begin{equation}
\begin{array}{l}
P=Q_{i}-Q_{o}=\left(Q_{h}+Q_{hl}\right)-\left(Q_{c}+Q_{hl}\right)
\end{array},\label{eq:Pot}
\end{equation}

\begin{equation}
\eta=1-\frac{Q_{c}+Q_{hl}}{Q_{h}+Q_{hl}},\label{eq:Efi}
\end{equation}
and
\begin{equation}
\Phi=T_{c}\sigma_{T}.\label{eq:Dis}
\end{equation}

Where $Q_{i}=Q_{h}+Q_{hl}$ and $Q_{o}=Q_{c}+Q_{hl}$ correspond to the input and output heat flows respectively. While $\sigma_{T}$ is the total entropy production of the heat engine plus its surroundings; that is, $\sigma_{T}=\sigma_{e}+\sigma_{w}$, where $\sigma_{e}$ is the entropy produced by the surroundings, given by:

\begin{equation}
\sigma_{e}=\frac{Q_{c}+Q_{hl}}{T_{c}}-\frac{Q_{h}+Q_{hl}}{T_{h}},\label{eq:EntroE}
\end{equation}
and $\sigma_{w}$ is the entropy produced by the heat engine,

\begin{equation}
\sigma_{w}=\frac{Q_{h}}{T_{h}}-\frac{Q_{c}}{T_{c}}+\sigma_{i}.\label{eq:EntroW}
\end{equation}

In the above equation, $\sigma_{i}$ represents the entropy production of the working substance due to different dissipative processes such as: turbulence, friction, viscosity, etc. This term represents the irreversibilities within the working substance. In the following sections, we will study two optimization cases of a type Curzon-Ahborn model taking into account two ways to quantify the irreversibilities. The first irreversible case is studied in terms of a lumped parameter $R$. This $R$ parameter, which comes from Clausius’s inequality, can be seen as a measure of the departure from the endoreversible regime. The second one includes the $r$-irreversible parameter that comes directly from the Clausius uncompensated heat. Hereinafter, we will name both of the studied cases as CA-$R$ case and CA-$r$ case, respectively. We point out the first irreversible case, without the heat leak term, does not reproduce the characteristic loops of real heat engines in the power output versus efficiency space. On the other hand, the second one reproduces the mentioned loops without the inclusion the heat leak term.

\subsection{CA-$R$ case with heat leak}
In this case, the parameter of non-reversibility $R$ quantifies the irreversibilities of the working substance and is related through a heat flux \cite{ParaR2,ParaR1} by,

\begin{equation}
\sigma_{i}=\left(1-R\right)\frac{Q_{c}}{T_{cw}},\label{eq:EntroSTNEndoRCC}
\end{equation}
as every thermal engine operating in cycles satisfies that $\sigma_{w}=0$, then by substituting Eq. \ref{eq:EntroSTNEndoRCC} into Eq. \ref{eq:EntroW} leads us for this model to the relationship between the reduced temperatures (high and low),

\begin{equation}
a_{c}\left(\gamma,R,a_{h}\right)=1-\frac{\gamma\left(1-a_{h}\right)}{Ra_{h}}.\label{eq:AcfAhNEHL}
\end{equation}
Thus, the heat flow $Q_ {c}$ can be rewritten as:

\begin{equation}
Q_{c}\left(\alpha,\gamma,T_{h},\tau,R,a_{h}\right)=\frac{T_{h}\alpha\tau\left(1-a_{h}\right)}{a_{h}\left(R+\gamma\right)-\gamma}.\label{eq:QcNEMHL}
\end{equation}

On the other hand, the process functions obtained by substituting the heat fluxes given by Eqs. \ref{eq:Qh}, \ref{eq:Qhl} and \ref{eq:QcNEMHL} into Eqs. \ref{eq:Pot}, \ref{eq:Efi} and \ref{eq:Dis} remain in function of $R$. In particular, for a heat engine, it is required that all the process functions are positive defined. To guarantee the above, it is necessary that the parameters $R$, $\delta$ and $a_{h}$ fulfilled with certain restrictions. Firstly, $a_{h}$ must be bounded by \cite{kPefi},

\begin{equation}
\frac{\gamma+\tau}{\gamma+R}<a_{h}<1.\label{eq:LahNEFC}
\end{equation}
Since in those points, the values of $P$ and $\eta$ are zero. Because of $a_h>0$ then $\tau<R\leq1$. In this CA-$R$ case, the parameters that allow to establish the configuration space \cite{Levario19,JEI2009} are essentially; the thermal conductance $\alpha$, the ratio conductances $\gamma$, as well as the relationship between the temperatures of the reservoirs ($\tau$). These parameters within the appropiate interval, controlling the energy flux that gets into the system. On the other hand, $\delta$ takes into accounts the amount of energy exchanged between external reservoirs restricting the performance of the converter. Finally, $a_{h}$ is associated with the operation modes. For this model, the process functions can be obtained by replacing Eqs. \ref{eq:Qh}, \ref{eq:Qhl} and \ref{eq:QcNEMHL} into Eqs. \ref{eq:Pot}, \ref{eq:Efi} and \ref{eq:Dis} obtaining respectively,

\begin{equation}
P=T_{h}\alpha\left(1-a_{h}\right)\left[1-\frac{\tau}{a_{h}\left(R+\gamma\right)-\gamma}\right],\label{eq:PotNEHL}
\end{equation}
\begin{equation}
\eta=\frac{\alpha\left(1-a_{h}\right)\left[\gamma+\tau-a_{h}\left(R+\gamma\right)\right]}{\left[\gamma-a_{h}\left(R+\gamma\right)\right]\left[\alpha\left(1-a_{h}\right)+\delta\left(1-\tau\right)\right]},\label{eq:EfiNEHL}
\end{equation}
 and
\begin{equation}
\Phi=T_{h}\left\{ \delta\left(1-\tau\right)^{2}-\frac{\alpha\tau\left(1-a_{h}\right)\left[1+\gamma-a_{h}\left(R+\gamma\right)\right]}{\gamma+a_{h}\left(R+\gamma\right)}\right\}. \label{eq:DisNEHL}
\end{equation}

Power output and efficiency have a value of high reduced temperature which maximizes them and therefore, this allows to characterize both the maximum power output and maximum efficiency regimes, respectively. The $a_h$ optimal value for $MP$ and $M\eta$ regimes are given by:

\begin{equation}
a_{h}^{MP}\left(\gamma,\tau,R\right)=\frac{\gamma+\sqrt{R\tau}}{R+\gamma}\label{eq:ahMaxPotNEHL}
\end{equation}
and
\begin{equation}
a_{h}^{M\eta}\left(\alpha,\delta,\gamma,\tau,R\right)=\frac{\left(R+\gamma\right)\left[\alpha\tau-\gamma\delta\left(1-\tau\right)\right]-\rho_{\eta R}}{\left(R+\gamma\right)\left[\alpha\tau-\delta\left(R+\gamma\right)\left(1-\tau\right)\right]},\label{eq:ahMaxEfiNEHL}
\end{equation}
with $\rho_{\eta R}$ of the form:
\begin{equation}
\rho_{\eta R}=\sqrt{R\tau\delta\left(R+\gamma\right)\left(1-\tau\right)\left[\gamma\delta\left(1-\tau\right)-\alpha\tau+R\left\lbrace\alpha+\delta\left(1-\tau\right)\right\rbrace\right]}.\label{eq:rEtaRNEHL}
\end{equation}

The mathematical expressions \ref{eq:ahMaxPotNEHL} and \ref{eq:ahMaxEfiNEHL} are the results of a derivative of power output and efficiency (obtained by substituting Eqs. \ref{eq:Qh}, \ref{eq:Qhl} and \ref{eq:QcNEMHL} into \ref{eq:Pot} and \ref{eq:Efi}) with respect to $a_h$, in order to find the reduced temperatures that cancel the derivatives. Maximum power output regime is completely defined within the CA-$R$ case ($\delta=0$). However, the heat leak term inclusion is needed to reach the maximum efficiency regime \cite{Andresen2001,Gordon3,kPefi}.

\subsection{CA-$r$ case with heat leak}
In this case, the parameter associated with internal irreversibilities is the Clausius uncompensated heat. So $\sigma_{i}$ can be expressed in the form \cite{SilvArias13,Clausius},

\begin{equation}
\sigma_{i}=r\alpha,\label{eq:EntroSTUCH}
\end{equation}
where $r$ is defined as the quotient between $\sigma_{S}$ and $\alpha$ which represents in some way the irreversibility degree of the working substance \cite{SilvArias13,LevarioArias19}.

For CA-$r$ case, the relationship between internal and external reduced temperatures is obtained by substituting Eq. \ref{eq:EntroSTUCH} into Eq. \ref{eq:EntroW}, due to that the working substance operates in cycles,

\begin{equation}
a_{c}\left(\alpha,\gamma,\tau,a_{h}\right)=1+\gamma\left(1-r\right)-\frac{\gamma}{a_{h}}
,\label{eq:AcfAhUCHHL}
\end{equation}
in such a way, the heat flux $Q_{c}$ is given by,

\begin{equation}
Q_{c}\left(\alpha,\delta,\gamma,\tau,T_{h},r,a_{h}\right)=\frac{T_{h}\alpha\tau\left[a_{h}\left(1-r\right)-1\right]}{\gamma-a_{h}\left[\gamma\left(1-r\right)+1\right]}.\label{eq:Q2UHCHL}
\end{equation}

Besides, the mathematical expressions for the process functions: power output ($P$), efficiency ($\eta$)  and dissipation ($\Phi$), can be obtained replacing Eqs. \ref{eq:Qh}, \ref{eq:Q2UHCHL} and \ref{eq:Qhl} into Eqs. \ref{eq:Pot}, \ref{eq:Efi} and \ref{eq:Dis}; that is,

\begin{equation}
P=T_{h}\alpha\left\{ 1-a_{h}+\frac{\tau\left[1-a_{h}\left(1-r\right)\right]}{\gamma-a_{h}\left[1+\gamma\left(1-r\right)\right]}\right\} ,\label{eq:PotIHL}
\end{equation}
\begin{equation}
\eta=\frac{\alpha\tau\left[1-a_{h}\left(1-r\right)\right]+\alpha\left( 1-a_{h}\right) \left\{ \gamma-a_{h}\left[1+\gamma\left(1-r\right)\right]\right\} }{\left[\delta\left(1-\tau\right)+\alpha\left(1-a_{h}\right)\right]\left[\gamma-a_{h}\left(1+\gamma\left[1-r\right]\right)\right]},\label{eq:EfiIHL}
\end{equation}
and
\begin{equation}
\Phi=T_{h}\frac{a_{h}^{2}\alpha\tau\left[1+\gamma\left(1-r\right)\right]-a_{h}\left\{\alpha\tau\left(1+\gamma\right)\left(2-r\right)-\delta\left(1-\tau\right)^{2}\left[1+\gamma\left(1-r\right)\right]\right\} +\alpha\tau+\gamma\left[\alpha\tau-\delta\left(1-\tau\right)^{2}\right]}{a_{h}\left[1+\gamma\left(1-r\right)\right]-\gamma}.\label{eq:DisIHL}
\end{equation}

As in the previous section, both power output and efficiency have a maximum, and both are zero at the same high reduced temperature values. While, dissipation function is a decreasing function with respect to the $a_h$ variable. Analogously, the zeros of $P$ and $\eta$ functions define the interval of values for $a_h$ that allow the energy converter to work as a heat engine,

\begin{equation}
    \frac{1+\gamma(2-r)+\tau(1-r)-n_{i}}{2[1+\gamma(1-r)]}\leq a_h\leq\frac{1+\gamma(2-r)+\tau(1-r)+n_{i}}{2[1+\gamma(1-r)]},\label{eq:limahIHL}
\end{equation}
where $n_i$ is given by,
\begin{equation}
    n_{i}=\sqrt{\left(1-\tau\right)^{2}-2r\left(1+\tau\right)\left(\gamma+\tau\right)+r^{2}\left(\gamma+\tau\right)^{2}},\label{eq:riIHL}
\end{equation}
because of $\tau$ is related to the thermal gradient that promotes the heat flow inside the converter and therefore, it limits the available energy in the system to access certain operation modes. Likewise, in order to $a_{h}$ satisfy the Ineq. \ref{eq:limahIHL}, the following condition must be fulfilled simultaneously \cite{kPefi}:

\begin{equation}
    0<r<\frac{\left(1-\sqrt{\tau}\right)^2}{\tau+\gamma}.\label{eq:limt1IHL}
\end{equation}
On the other hand, it is always desirable that $\delta$ to be lower than the $\alpha$ value, so that a greater entering heat flux to the system is guaranteed, i.e, a inequality must be kept $0\leq\delta<\alpha$.

In this scheme, as in the CA-$R$ case, there are high reduced temperature values that allow us to obtain both the maximum power output and maximum efficiency regimes. These reduced temperatures are given by,

\begin{equation}
a_{h}^{MP}\left(\gamma,\tau,r\right)=\frac{\gamma+\sqrt{\tau}}{1+\gamma\left(1-r\right)},\label{eq:ahMaxPIHL}
\end{equation}
and
\begin{equation}
a_{h}^{M\eta}\left(\alpha,\delta,\gamma,\tau,r\right)=\frac{\alpha\tau+\gamma\left\lbrace\alpha\tau\left(1-r\right)-\delta\left(1-\tau\right)\left[1+\gamma\left(1-r\right)\right]\right\rbrace-\rho_{\eta r}}{\left[1+\gamma\left(1-r\right)\right]\left\lbrace\alpha\tau\left(1-r\right)+\delta\left(1-\tau\right)\left[1+\gamma\left(1-r\right)\right]\right\rbrace},\label{eq:ahMaxEfiIHL}
\end{equation}
with $\rho_{\eta r}$ given by:
\begin{equation}
\rho_{\eta r}=\sqrt{\tau\left[1+\gamma\left(1-r\right)\right]\left\lbrace\delta^{2}\left(1-\tau\right)^{2}\left[1+\gamma\left(1-r\right)\right]+r\alpha^{2}\tau+\alpha\delta\left(1-\tau\right)\left[1-r\left(\gamma-\tau\right)\right]-\tau\right\rbrace}.\label{eq:rEtarIHL}
\end{equation}

The $a_h$ optimal values given by Eqs. \ref{eq:ahMaxPIHL} and \ref{eq:ahMaxEfiIHL} are obtained by replacing Eqs. \ref{eq:Qh}, \ref{eq:Qhl} and \ref{eq:Q2UHCHL} into \ref{eq:Pot} and \ref{eq:Efi}. In this case, it is important to note that the maximum efficiency regime exists without the need to consider a heat leak (see Fig. \ref{fig:PPvsE} c). In the following section, we establish the necessary conditions for the parametric curves ($P$ vs $\eta$ and $P$ vs $\Phi$) in order to they are compatible with the characteristic operation mode.

\section{Configuration and energetic reconfiguration of some power plants}

\begin{figure}
	\centering
		\includegraphics[scale=.75]{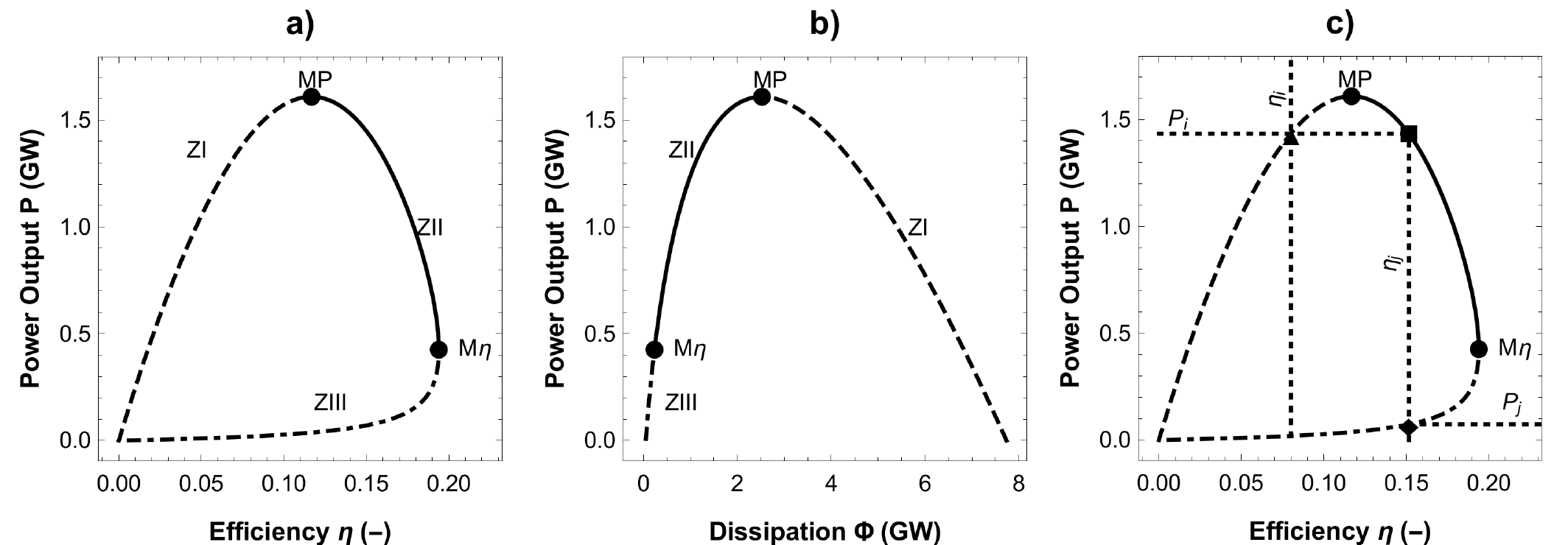}
	\caption{a) Parametric curves of power output vs efficiency and b) power output vs dissipation, which show the operating characteristic zones for a heat engine. In c), the parametric graph of power output vs efficiency shows how an energy converter can achieve a power output value $P_i$ with two different efficiency values ($\eta_i$ and $\eta_j$) at the same time, an efficiency value ($\eta_j$) has associated two different power output values ($P_i$ and $P_j$).}
	\label{fig:PPvsE}
\end{figure}

During the operation of several thermal engines a great number of parameters associated with the exchange of energy are taken into account, all of them give rise to a configuration space. In this space, there is an infinity of compatible operation modes with the performance of every energy converter \cite{Andresen2001,JEI2009}. This space is formed by a unique combination of phenomenological parameters, and they are related to each other through every converter model. The used irreversible model in this work incorporate the most representative elements during the energy transfer, such as the thermal conductances ($\alpha$, $\delta$ and $\beta$), as well as the $\gamma$ parameter \cite{Levario19,JEI2009}. Other important parameters are the temperatures ratio of the external reservoirs ($\tau$), related to the capacity of the system to promote an effective heat flow and the parameters associated with the way to quantify the internal irreversibility degree ($R$, $r$).

When the behavior of the $P$ vs $\eta$ curves in the CA-$R$ and CA-$r$ cases is analyzed, a particular curve (loop) can be observed and three operating zones are well distinguished (see Figs. \ref{fig:PPvsE}). In each zone, there are specific operation modes that allow the energy converter to achieve a unique performance. The operation modes located in zone $I$ ($ZI$) are usually characterized by an $a_{h}\in\left(a_{h0},a_{h}^{MP}\right)$, where $a_{h0}$ is the value of the high reduced temperature from which the converter starts to work as a heat engine and $a_{h}^{MP}$ value corresponds to the maximum power output regime, these operation modes have high dissipation ($HD$) and low efficiency ($LE$). The operation of power plants in zone $II$ ($ZII$) is distinguished by $a_{h}\in\left[a_{h}^{MP},a_{h}^{M\eta}\right]$ where $a_{h}^{M\eta}$ is the high reduced temperature value of the maximum efficiency regime. In this zone, the converter performance reaches high power output ($HP$), good efficiency ($HE$) and a moderate dissipation. On the other hand, when the operation is performed in zone $III$ ($ZIII$), we get an $a_{h}\in\left(a_{h}^{M\eta},a_{h1}\right)$, where $a_{h1}$ is the upper bound for the high reduced temperature that restricts every energy converter to operate as a heat engine. The operation modes in $ZIII$ have low dissipation and low power output ($LD$ and $LP$). It is understood by $HE$ an efficiency greater than the efficiency of maximum power output regime. In $HD$, the highest dissipation values remain bounded by the maximum power output regime, while $LD$ refers to dissipation values lower than the dissipation at maximum efficiency regime. Every power output value greater than the power output at maximum efficiency regime can be considered in $HP$, all of the above is shown in Figs. \ref{fig:PPvsE}a and \ref{fig:PPvsE}b. Another particularity in Fig. \ref{fig:PPvsE}c is observed, an energy converter can have the same power output value for two different efficiency values. Similarly, for a given efficiency value, the converter can develop two completely different power output values.

In Table \ref{tab:DataPlant} some reported operating data are shown for some power plants (temperatures of the reservoirs, power output and efficiency values) and by using any of the CA cases, it is possible to mark off with greater precision the values in the configuration space that guarantees a power plant to operate as a heat engine, and simultaneously give the necessary conditions to study the quality in its operation.

\begin{table}
\begin{centering}
\begin{tabular}{|c|c|c|c|c|c|}
\hline 
\multicolumn{2}{|c|}{{\footnotesize{}Almaraz II (A)}} & \multicolumn{2}{c|}{{\footnotesize{}West Thurrock (WT)}} & \multicolumn{2}{c|}{{\footnotesize{}Toshiba (T)}}\tabularnewline
\multicolumn{2}{|c|}{{\footnotesize{}(PWR, Spain, 83)}} & \multicolumn{2}{c|}{{\footnotesize{}(Uk, 62)}} & \multicolumn{2}{c|}{{\footnotesize{}(109FA, 04)}}\tabularnewline
\hline 
{\footnotesize{}$T_{h}[K]$} & {\footnotesize{}$T_{c}[K]$} & {\footnotesize{}$T_{h}[K]$} & {\footnotesize{}$T_{c}[K]$} & {\footnotesize{}$T_{h}[K]$} & {\footnotesize{}$T_{c}[K]$}\tabularnewline
\hline 
{\footnotesize{}600} & {\footnotesize{}290} & {\footnotesize{}838} & {\footnotesize{}298} & {\footnotesize{}1573} & {\footnotesize{}303}\tabularnewline
\hline 
{\footnotesize{}$P[GW]$} & {\footnotesize{}$\eta[-]$} & {\footnotesize{}$P[GW]$} & {\footnotesize{}$\eta[-]$} & {\footnotesize{}$P[GW]$} & {\footnotesize{}$\eta[-]$}\tabularnewline
\hline 
{\footnotesize{}1.044} & {\footnotesize{}0.35} & {\footnotesize{}1.240} & {\footnotesize{}0.36} & {\footnotesize{}0.342} & {\footnotesize{}0.48}\tabularnewline
\hline 
\multicolumn{2}{|c|}{{\footnotesize{}Cofrentes (C)}} & \multicolumn{2}{c|}{{\footnotesize{}Lardarello (L)}} & \multicolumn{2}{c|}{{\footnotesize{}Alstom (Al)}}\tabularnewline
\multicolumn{2}{|c|}{{\footnotesize{}(BWR, Spain, 84)}} & \multicolumn{2}{c|}{{\footnotesize{}(Italy,64)}} & \multicolumn{2}{c|}{{\footnotesize{}(ka26-1)}}\tabularnewline
\hline 
{\footnotesize{}$T_{h}[K]$} & {\footnotesize{}$T_{c}[K]$} & {\footnotesize{}$T_{h}[K]$} & {\footnotesize{}$T_{c}[K]$} & {\footnotesize{}$T_{h}[K]$} & {\footnotesize{}$T_{c}[K]$}\tabularnewline
\hline 
{\footnotesize{}562} & {\footnotesize{}289} & {\footnotesize{}523} & {\footnotesize{}353} & {\footnotesize{}1398} & {\footnotesize{}288}\tabularnewline
\hline 
{\footnotesize{}$P[GW]$} & {\footnotesize{}$\eta[-]$} & {\footnotesize{}$P[GW]$} & {\footnotesize{}$\eta[-]$} & {\footnotesize{}$P[GW]$} & {\footnotesize{}$\eta[-]$}\tabularnewline
\hline 
{\footnotesize{}1.092} & {\footnotesize{}0.34} & {\footnotesize{}0.150} & {\footnotesize{}0.16} & {\footnotesize{}0.410} & {\footnotesize{}0.57}\tabularnewline
\hline 
\end{tabular}
\par\end{centering}
\caption{\label{tab:DataPlant}Operating reported data of some power plants: temperatures of energy reservoirs ($T_{h}$ and $T_{c}$), power output (P) and efficiency ($\eta$). Two nuclear (Almaraz II and Cofrentes plants), two mono-cycle (West Thurrock and Larderello plants) and two of combined cycle (Toshiba and Alstom plants).}
\end{table}

\subsection{Energetic configuration of CA-$R$ and CA-$r$ cases}
When energetic performance of different power plants under CA-$R$ and CA-$r$ cases is analyzed, it is needed more information than the provided in Table \ref{tab:DataPlant} to build a complete configuration space. Where efficiency and power output are represented by a point in the configuration space, there is a great number of curves and each of them represents a particular combination of construction parameters, that allow the converter to reach specific values of $P$ and $\eta$. There are several non reported data during the performance but is possible to include them by using the operating bounds imposed by each case.

\subsubsection{CA-$R$ case}

\begin{figure}
	\centering
		\includegraphics[scale=.75]{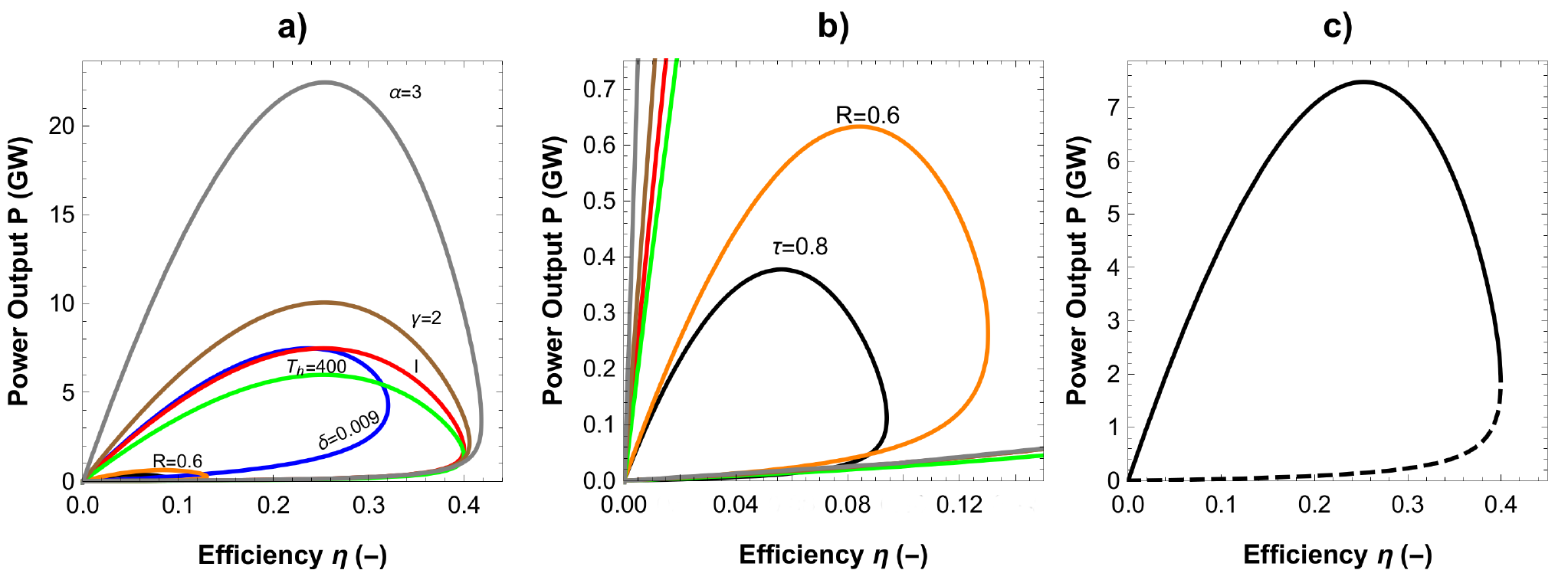}
	\caption{In a) and b), curves of power output vs efficiency for a CA-$R$ case with heat leak and different configurations, in each of them some parameters have been varied, on the basis of (I) $\alpha=1$, $\gamma=3$, $\tau=0.5$, $\delta=0.001$, $R=0.9$ y $T_{h}=500 K$ are presented. In c) we show how $+$ sign (solid line) and $-$ sign (dashed line) of Eq. \ref{eq:ParaPEtaR}, describe the operation modes of curve (I).}
	\label{fig:PvsENEHL}
\end{figure}

In this case, we determine the parameters $\delta$, $R$ and $a_{h}$ require extra--thermodynamic constraints to fully characterize a compatible point with an operation mode in the configuration space. As $R$ quantifies to some degree the internal irreversibilities within the working substance during the operation of a heat engine, it can also be associated with the energy dissipated by operation cycle. $\delta$-parameter modulates the amount of energy that leaks from the system and therefore plays no role in the operation of the converter. With the inclusion of these irreversibilities, it is guaranteed the parametric curves $P$ vs $\eta$ form the well--known loop. Figs. \ref{fig:PvsENEHL}a and \ref{fig:PvsENEHL}b reflect the use of $a_{h}(\eta)$ as the parametric variable. Since it is always possible to isolate $a_h$ from the expression for efficiency (Eq. \ref{eq:EfiNEHL}). The parametric variable $a_{h}(\eta)$ is substituted in the expression for power output (Eq. \ref{eq:PotNEHL}) and we have two equations in terms of efficiency, given by

\begin{equation}
   P_{CA-R}= T_{h}\eta \frac{R[\alpha(1-\eta)+\delta(2-\eta)(1-\tau)]+\gamma\delta(2-\eta)(1-\tau)-\alpha\tau\pm \rho_{pR}}{2\alpha(R+\gamma)(1-\eta)},\label{eq:ParaPEtaR}
\end{equation}
With $\rho_{pR}$,

\begin{equation}
    \rho_{pR}=\sqrt{[\gamma\delta\eta+R\eta(\alpha+\delta)-R\alpha]^2-2\tau\left\lbrace R\alpha^2-\alpha\eta\left[R\alpha-\delta\left(1-R\right)\left(R+\gamma\right)\right]+\delta\eta^2(R+\gamma)\left[\gamma\delta+R\left(\alpha+\delta\right)\right]\right\rbrace+d_{R}}.\label{eq:rpR}
\end{equation}
and
\begin{equation}
    d_{R}=\tau^2[\alpha+\delta\tau(R+\gamma)]^2.
\end{equation}
From Eq. \ref{eq:ParaPEtaR} $+$ sign describes the points located in zones $I$ and $II$, while $-$ sign allows to plot the points situated in zone $III$ (as shown in Fig. \ref{fig:PvsENEHL}c).

The CA-$R$ case allows us to get some relationships between the parameters related to the main sources of irreversibilities ($\delta$ and $R$) and the variables that modulate the energy input to the system ($\alpha$, $\gamma$ and $T_{h}$). The mathematical expressions for the process functions ($P$, $\eta$ and $\Phi$) can modify the performance of a converter. In Figs. \ref{fig:PvsENEHL}a and \ref{fig:PvsENEHL}b is also observed how the curves ($P$ vs $\eta$) change when at less one of the parameters varies. For instance, if $\alpha$ or $\gamma$ shifts power output has a variation, without considerably reducing the efficiency. Whether $T_{h}$ is modified, the effect is only reflected in the power output. On the contrary, when the parameters $\tau$, $\delta$ or $R$ change, both $P$ and $\eta$ are significantly altered (see blue loop in Fig. \ref{fig:PvsENEHL}a with $\delta=0.009$ and orange loop with $R=0.6$ in Fig. \ref{fig:PvsENEHL}b). 

To make evident all of the above, a direct relationship between $R$ and $\delta$ with $P$ and $\eta$ is established. With the Eq. \ref{eq:ParaPEtaR}, it is guaranteed a specific curve contains the corresponding operation mode with the power and efficiency reported by the analyzed plants. Moreover, if $P\geq0$ (Eq. \ref{eq:ParaPEtaR}) for any operation zone, the parameter $\delta$ has two possible expressions:

\begin{equation}
\delta_{(ZI),(ZII,III)}=\frac{P\left(R+\gamma\right)\left(2-\eta\right)-\eta\left[T_{h}\alpha\left(R-\text{\ensuremath{\tau}}\right)\mp\sqrt{P^{2}\left(R+\gamma\right)^{2}+T_{h}^{2}\alpha^{2}\left(R-\tau\right)^{2}-2PT_{h}\alpha\left(R+\gamma\right)\left(R+\tau\right)}\right]}{2T_{h}\eta\left(R+\gamma\right)\left(1-\tau\right)}\label{eq:DeInZ}
\end{equation}

The value of $\delta_{(ZI)}$, given by the $-$ sign, associates the reported operation mode by any of the power plants with a particular energy configuration in $ZI$. Therefore, $\alpha$-parameter must be bounded as follows:

\begin{equation}
\frac{P\left(R+\gamma\right)\left(\sqrt{R}+\sqrt{\tau}\right)^{2}}{T_{h}\left(R-\tau\right)^{2}}\leq\alpha_{ZI}<\frac{P\left(R+\gamma\right)\left(1-\eta\right)}{T_{h}\eta\left[R\left(1-\eta\right)-\tau\right]},\label{eq:lAlZ1NEHL}
\end{equation}
with this condition, it is guaranteed that $\delta$ is non-negative. Ineq. \ref{eq:lAlZ1NEHL} is positive when the parameter $R$ fulfill,

\begin{equation}
\frac{\tau}{\left(1-\eta\right)^{2}}<R\leq1.\label{eq:LRZ1NEHL}
\end{equation}

Likewise, to be physically consistent this condition must be satisfied:

\begin{equation}
0<\tau<\left(1-\eta\right)^{2}.\label{eq:lTauZ1NEHL}
\end{equation}

For the possible $\delta_{(ZII,III)}$ values provided by Eq. \ref{eq:DeInZ} ($+$ sign), which relate a characteristic operation regime to a configuration in $ZII$ or $ZIII$ ($\delta_{(ZII,III)}>0$), then $\alpha$ parameter must be,

\begin{equation}
    \alpha_{(ZII,III)}>\frac{P\left(R+\gamma\right)\left(1-\eta\right)}{T_{h}\eta\left[R\left(1-\eta\right)-\tau\right]},\label{eq:CAlp23NEHL}
\end{equation}
since these $\alpha$ values make the parameter $\delta_{(ZII,III)}$ equal to zero. In addition, to ensure that Eq. \ref{eq:CAlp23NEHL} must be positive, the $R$ parameter will be bounded by,

\begin{equation}
\frac{\tau}{1-\eta}<R\leq1,\label{eq:lRZ23NEHL}
\end{equation}
we note the lower bound of $R$ is an asymptotic point, this makes $\alpha$ tends to infinity. Therefore, $R$ must satisfy the Ineq. \ref{eq:lRZ23NEHL} and so,

\begin{equation}
0<\tau\leq\left(1-\eta\right).\label{eq:lTauZ23NEHL}
\end{equation}

Although Eq. \ref{eq:lTauZ1NEHL} is contained in Eq. \ref{eq:lTauZ23NEHL}, when the relationship between $\tau$ and $\eta$ satisfy exclusively the constraint in Eq. \ref{eq:lTauZ1NEHL}, some configurations show up for reported operation modes in $ZI$.

\subsubsection{CA-$r$ case}

\begin{figure}
	\centering
		\includegraphics[scale=0.8]{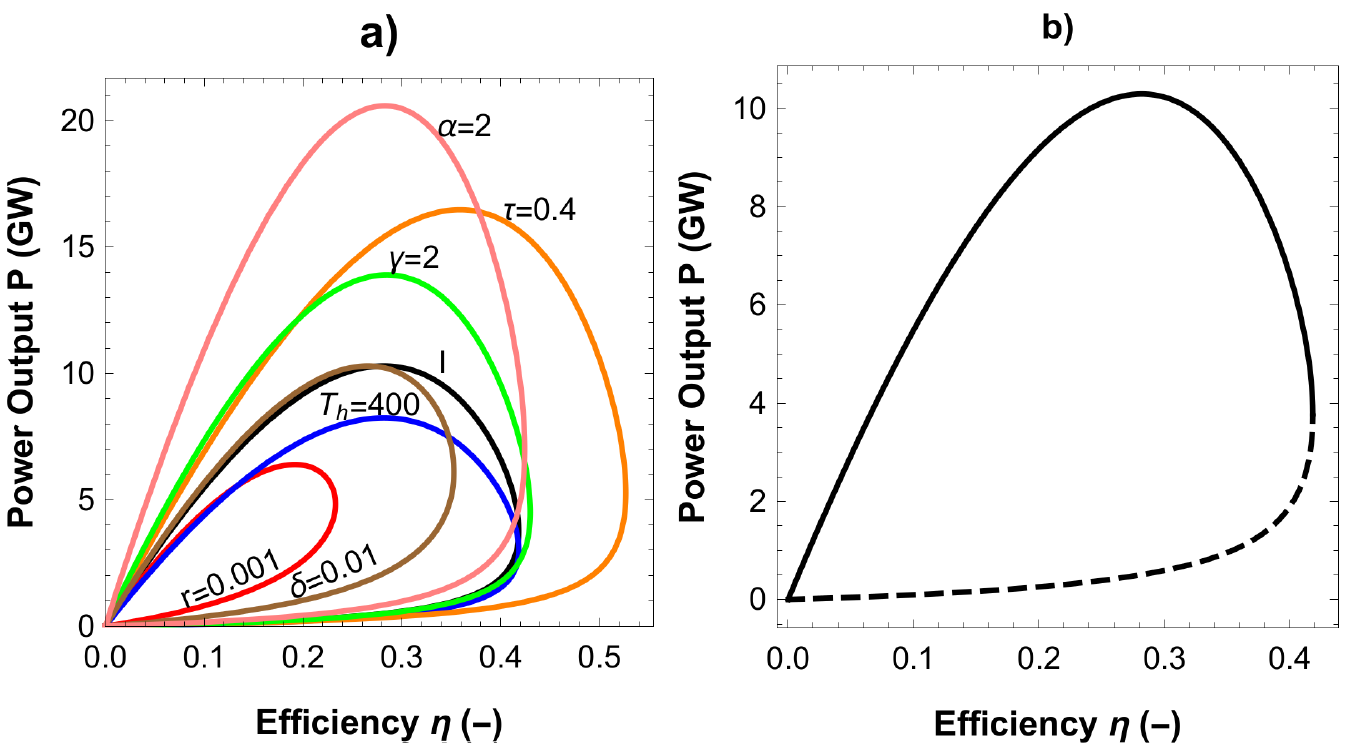}
	\caption{In a), several configurations for power output vs efficiency plane of the CA-$r$ case. In each configuration the model's parameters have been considered, taking as reference the loop $I$, built with the values of: $\alpha=1$, $\delta=0.001$, $\gamma=3$, $\tau=0.5$, $T_{h}=500$ and $r=0.001$. In b), loop $I$ is shown with its characteristics zones described by $+$ sign (solid line) and $-$ sign (dashed line) of Eq. \ref{eq:PrZ12IHL}.}
	\label{fig:PvsEIHL}
\end{figure}
In this model, we define once again the constraints for the parameters $a_{h}$, $r$ and $\delta$ in such a way they allow the construction of a curve in the configuration space, and compatible with a particular operation model. In this case, $r$ reflects the internal irreversibilities of the system and therefore, is associated with the dissipated energy during the operation of the heat engine. Although $r$ is the only parameter that allows the model to characterize the loops in the $P$ vs $\eta$ plane, $\delta$-parameter affects the maximum value that the efficiency can reach (Fig. \ref{fig:PvsEIHL} a). Hence, some restrictions are imposed on the $\delta$-parameter. They are directly related to the parametric equation ($P=P(\eta)$):

\begin{equation}
P_{CA-r}=T_{h}\eta\frac{\alpha\left(1-r\gamma\right)\left(1-\eta\right)+\delta\left(1-\tau\right)\left(2-\eta\right)\left[1+\gamma\left(1-r\right)\right]-\alpha\tau\left(1-r\right)\mp \rho_{1r}}{2\left(1-\eta\right)\left[1+\gamma\left(1-r\right)\right]},\label{eq:PrZ12IHL}
\end{equation}
where $-$ sign locates the operation modes in $I$ and $II$ zones. While in Zone $III$ the modes are described by $+$ sign (see Fig. \ref{fig:PvsEIHL} b). Thus, $\rho_{1r}$ is given by,

\begin{equation}
\rho_{1r}=\sqrt{\left\{\delta\eta\left(1-\tau\right)\left[1+\gamma\left(1-r\right)\right]-\alpha\left(1-\eta\right)\left[1+\gamma\left(2-r\right)\right]-\alpha\tau\left(1-r\right)\right\}^{2}+d_{1r}},\label{eq:r1r}
\end{equation}
with $d_{1r}$
\begin{equation}
d_{1r}=4\alpha\left(1-\eta\right)\left[1+\gamma\left(1-r\right)\right]\left[\gamma\delta\eta-\alpha\gamma\left(1-\eta\right)-\tau\left(\alpha+\delta\tau\eta\right)\right].\label{eq:d1r}
\end{equation}

Analogous to the CA-$R$ case, we found a curve (loop) compatible with the reported efficiency and power output values in the power plants data. The $\delta$ parameter has two cases according to the operation zone,

\begin{equation}
\delta_{(ZI),(ZII,III)}=\frac{P\left(2-\eta\right)\left[1+\gamma\left(1-r\right)\right]+\eta\left\lbrace T_{h}\alpha\left[\tau+r\left(\gamma-\tau\right)-1\right]\mp \rho_{\delta}\right\rbrace}{2T_{h}\left(1-\tau\right)\left[1+\gamma\left(1-r\right)\right]}\label{eq:DcP}
\end{equation}
where $-$ sign refers to $ZI$ and $+$ sign corresponds to $ZII$ and $ZIII$, besides:

\begin{equation}
\rho_{\delta}=\sqrt{\left\lbrace P\left[1+\gamma\left(1-r\right)\right]+T_{h}\alpha\left(r\gamma-1\right)\right\rbrace^{2}-2T_{h}\alpha\tau\left\lbrace T_{h}\alpha+\left[P\left(1-r\right)+rT_{h}\alpha\right]\left[1+\gamma\left(1-r\right)\right]\right\rbrace+T_{h}^{2}\alpha^{2}\tau^{2}\left(1-r\right)^{2}}.\label{eq:rdihl}
\end{equation}
To ensure that $\delta_{ZI}$ and $\delta_{(ZII,III)}$ are positive, $\alpha$ and $r$ parameters must be constrained. All of the operation modes in $ZI$ must fulfill:

\begin{equation}
P\left[\frac{1}{P-T_{h}\alpha\eta}+\frac{1-\eta}{P\gamma\left(1-\eta\right)+T_{h}\alpha\eta\tau}\right]<r<\frac{P\left(1+\gamma\right)-T_{h}\alpha\left(1-\sqrt{\tau}\right)^{2}}{P\gamma-T_{h}\alpha\left(\gamma+\tau\right)},\label{eq:rAIHL1}
\end{equation}
this condition guarantees that $\delta_{Z1}>0$. As long as,

\begin{equation}
\frac{P\left(1+\gamma\right)}{T_{h}\left(1-\sqrt{\tau}\right)^{2}}<\alpha_{ZI}<\frac{P\left[\gamma\left(1-\eta\right)+\sqrt{\tau}\right]}{T_{h}\eta\sqrt{\tau}\left(1-\sqrt{\tau}\right)},\label{eq:lAIHL1}
\end{equation}
and at the same time $\alpha_{Z1}>0$. In the same way, for every operation mode in $ZII$ or $ZIII$, since $\delta_{Z2,3}>0$ then by transitivity:

\begin{equation}
0<r<P\left[\frac{1}{P-T_{h}\alpha\eta}+\frac{1-\eta}{P\gamma\left(1-\eta\right)+T_{h}\alpha\eta\tau}\right],\label{eq:rAIHL23} 
\end{equation}
and
\begin{equation}
\frac{P\left(1+\gamma\right)\left(1-\eta\right)}{T_{h}\eta\left[1-\left(\eta+\tau\right)\right]}<\alpha_{ZII,III}.\label{eq:lAIHL23}
\end{equation}

In the CA-$R$ and CA-$r$ cases, three particular operation zones are characterized. Although there are thermal engines that can operate in $ZI$ or $ZIII$, it is desirable they operate within $ZII$. In the following section, we will show that from a specific parametric variable, it is possible to find conditions to project operation modes outside of $ZII$ to modes within it.

\subsection{Energetic reconfiguration of CA-$R$ and CA-$r$ cases}
\begin{figure}
	\centering
		\includegraphics[scale=.8]{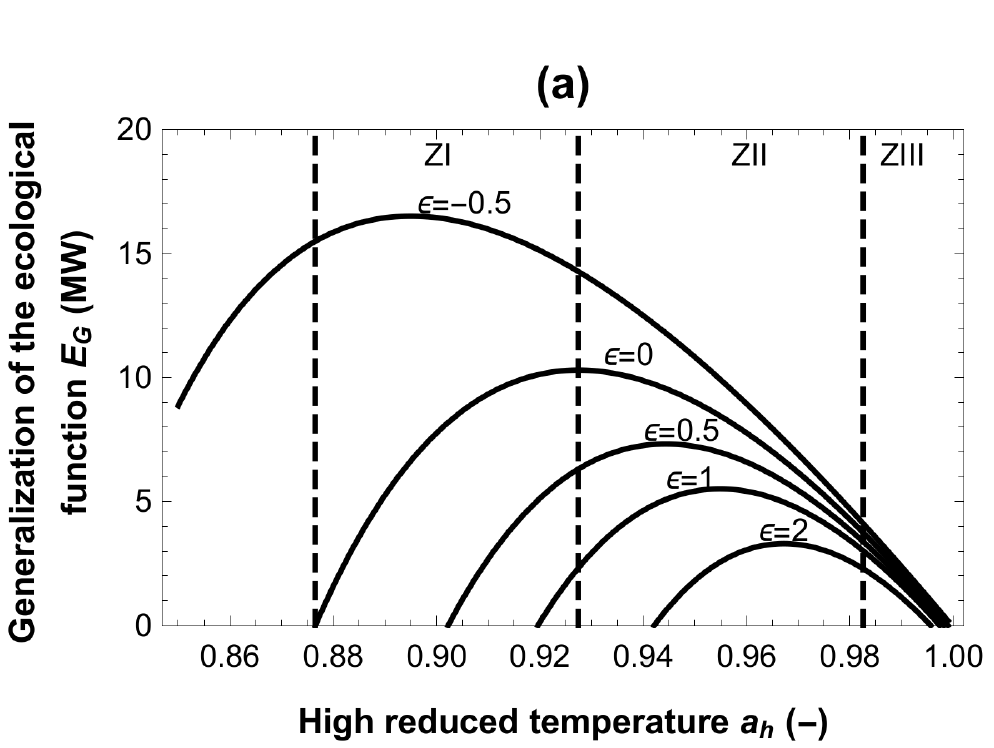}\includegraphics[scale=.73]{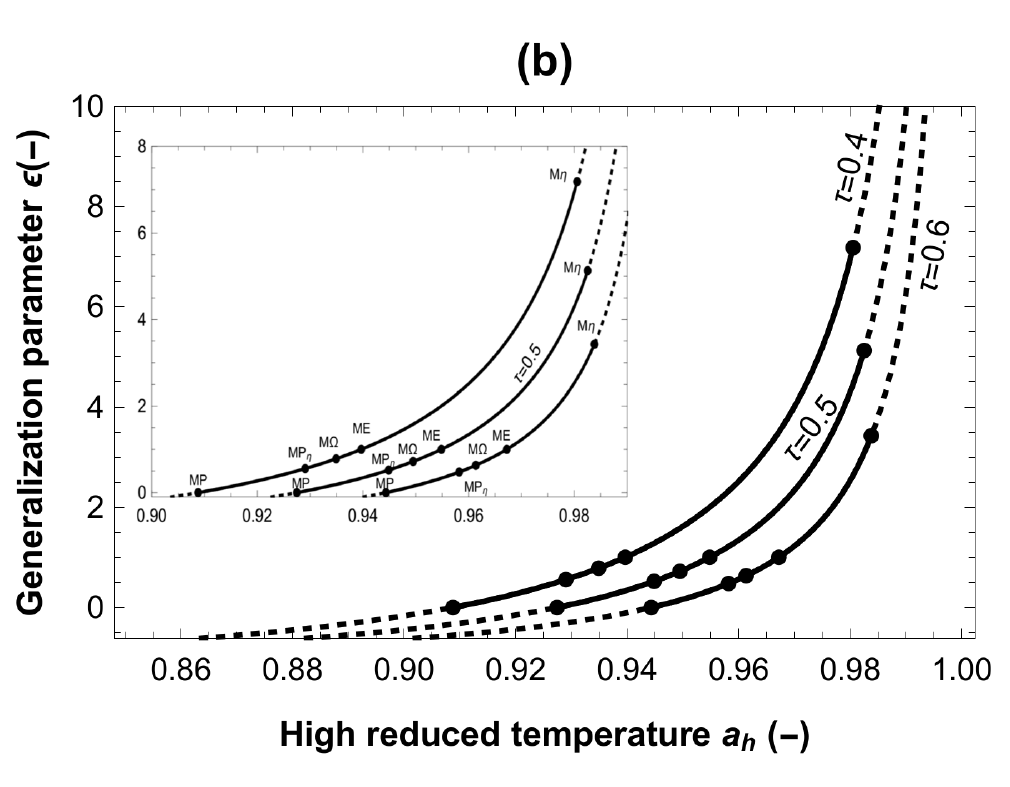}
	\caption{a) Generalization of the ecological function for $\tau=0.5$, $\gamma=3$ and several values of $\epsilon$. We observe that each of its maxima are in a respective operation zone. b) generalization parameter ($\epsilon$) for different values of $\tau$. In these graphs we observe how the parameter $\epsilon$ is linked to each $a_h$ value that maximize the some objective functions, maximum power output ($MP$), maximum efficient power ($MP_\eta$), maximum omega function ($M\Omega$) and maximum ecological function ($ME$).}
	\label{fig:G0new}
\end{figure}

Within the context of the FTT it has been possible to define certain objective functions, this type of functions are connected to the performance of an energy converter with a particular operation mode. Power output \cite{CAMod,Salamon01} and efficiency functions are the immediate examples \cite{Ocampo18,Schwalbe18}. Other objective functions have been used in the optimization of endoreversible heat engines such as omega function ($\Omega$ \cite{Calvo01}), efficient power ($P_{\eta}$ \cite{Yilmaz}) and the ecological function ($E$).  $E$ function was introduced by Angulo--Brown in 1991 \cite{Angulo91}, which is defined as $E=P-T_{c}\sigma_T$; where $P$ is the produced power output, $T_c$ is the temperature of the cold reservoir and $(\sigma_T)$ the total entropy production. The maximization of the ecological function leads to an engine configuration with a power output around $75\%$ of the maximum power output, and an entropy production around $25\%$ of the entropy produced in the maximum power regime. This property is called in the literature as the corollary $75-25$ \cite{GFE1}. Another ecological function property is that the efficiency that maximizes $E$-function is almost the semi-sum of the Curzon--Ahlborn ($\eta_{CA}$) and the Carnot ($\eta_{C}$) efficiencies \cite{Angulo91}. The ecological function has been widely applied in the context of FTT for several energy converter models; for instance, thermal, chemical and electrical engines \cite{Chen10,Ocampo18,Ocampo20}, as well as, biochemical reactions \cite{Bioc1,Bioc2} and atmospheric convective cells \cite{Barranco1996,Barranco2003}. Later in 2001, Angulo--Brown and Arias--Hern\'{a}ndez showed that a more suitable ecological function depends on the used heat transfer law to model the irreversible heat fluxes between the heat reservoirs and the working substance. For example, in the case of a Newtonian heat transfer law, the ecological function that gives the best compromise between high power output an low dissipation is given by $E=P-\sqrt{T_{h}T_{c}}\sigma$. Thus, depending on the heat transfer law used in the model, the generalized ecological function is always written as:

\begin{equation}
E_{G}=P-\epsilon\Phi,\label{eq:DefEcoGen}
\end{equation}
this $\epsilon$-parameter generates a family of convex ecological functions. Each of them with a value of $a_h$ that maximizes themselves (see Fig. \ref{fig:G0new} a). In particular, for the two analyzed irreversible cases in this work, such $a_h$'s depends on $\epsilon$, $\gamma$, $\tau$ and $R$ or $r$ parameters. Each generated ecological function represents a trade off between power output and dissipation. These trade offs are linked to a particular operation mode, which can be characterized by a value of $a_h$ or $\epsilon$, some operation modes correspond to other well-known objective functions (see Fig. \ref{fig:G0new} b). Recently in \cite{kPefi}, it has shown that a Newtonian heat transfer law allow us to link a generalization parameter with any operation mode. 

Up to this point, we have given guidelines so that different physically reachable operation modes are located on a loop described by a particular configuration space. However, there are operation modes that are not in $ZII$ and can be led to this one, by considering the following possibilities: maintaining the reported power output and improving the efficiency, retaining the reported efficiency and raising the power output or simply finding a new configuration where the reported power output and efficiency values are in $ZII$. To achieve these improvements, we use the ecological function's generalization parameter \cite{GFE1} and thus, we explore other operation modes to which the power plants of the Table \ref{tab:DataPlant} could access. All of the above allows us to establish the so-called ''improvement condition'' \cite{Levario19,kPefi} and through the maximum efficiency regime, we find other conditions with which each power plant would be operating in $ZII$.

\subsubsection{CA-$R$ case}
\begin{figure}
	\centering
		\includegraphics[scale=.55]{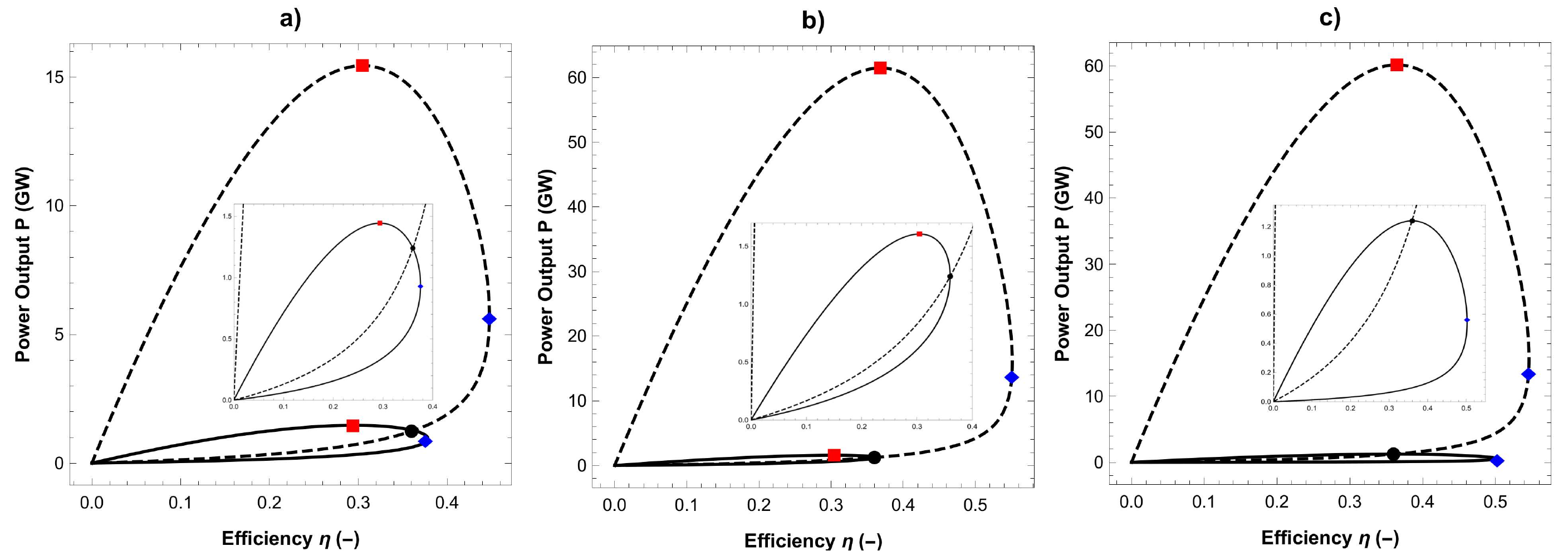}
	\caption{Characteristics loops for West Thurrock plant. In a) the loops exemplify two configurations, one of them lies in $ZII$ (solid line). The other one (dotted line) represents a configuration whose operation mode is in $ZIII$. In b) it is shown a configuration (solid line) in which, the reported mode for WT plant represents the maximum efficiency. In dotted line, the same operation mode is located in $ZIII$ with new features. In c) a specific configuration is sketched, in which the operation mode under new conditions for WT plant represents the maximum efficiency point (solid line). The other configuration contains the same operation mode is in $ZIII$ (dotted line) with new features.}
	\label{fig:G04}
\end{figure}

The purpose of energetic restructuring lies in finding conditions of certain elements (heat exchangers) for different power plants, so that they operate in $ZII$. In Fig.\ref{fig:G04} a, two different configurations are shown, for which West Thurrock plant (WT) can operate (see Tab. 1). In the first one, the operation mode with $\alpha=0.95 GW/K$, $\gamma=3$, $R=0.75$ and $\delta=1.951 MW/K$, is in $ZIII$. While the other one, whose configuration is $\alpha=0.078 GW/K$, $\gamma=3$, $R=0.784$ and $\delta=910.525 W/K$, it is shown that its associated operation mode lies in $ZII$. In Fig. \ref{fig:G04}b, for the reported data in Tab. \ref{tab:DataPlant}, a configuration whose design parameters are: $\alpha=67.941 MW/K$, $\delta=1.75 MW/K$, $\gamma=3.5$ and $R=0.9$ (loop in solid line), the operation mode emulates the maximum efficiency regime. In Fig. \ref{fig:G04}c, the maximum power output regime is obtained when the construction parameters are: $\alpha=0.052GW/K$, $\delta=196.02W/K$, $\gamma=3.5$ and $R=0.9$,  also in solid line. When those cases are compared, the energetic performance of this type of thermal engines can improve (better power output and efficiency), if some control parameters are varied. In particular, for operation modes in $ZII$ or $ZIII$, it is necessary to have well defined the point that corresponds to the regime of maximum efficiency, because it separates both zones. Replacing Eq. \ref{eq:ahMaxEfiNEHL} in the mathematical expression for the power output (Eq. \ref{eq:Pot}) and after solving for $\alpha$-parameter, the new solution helps us to find the first transition condition between zones $ZIII\rightarrow ZII$ is,

\begin{equation}
\alpha=\frac{\delta\eta\left(R+\gamma\right)\left(1-\tau\right)\left[R\left(1-\eta\right)+\tau+2\sqrt{R\tau\left(1-\eta\right)}\right]}{\left[\tau-R\left(1-\eta\right)\right]^{2}}.\label{eq:AlfMaxEfiNEHL}
\end{equation}

The other conditions arise from evaluating the efficiency function at the point of maximum efficiency (replacing Eq. \ref{eq:ahMaxEfiNEHL} into Eq. \ref{eq:Efi}) and after solving for $\delta$ we get,

\begin{equation}
\delta=\alpha\frac{P\left(R+\gamma\right)\left(R+\tau\right)+\left(R-\tau\right)\left[\sqrt{P^{2}\left(R+\gamma\right)^{2}+T_{h}^{2}\alpha^{2}\left(R-\tau\right)^{2}-2PT_{h}\alpha\left(R+\gamma\right)\left(R+\tau\right)}-T_{h}\alpha\left(R-\tau\right)\right]}{2\left(R+\gamma\right)\left(1-\tau\right)\sqrt{P^{2}\left(R+\gamma\right)^{2}+T_{h}^{2}\alpha^{2}\left(R-\tau\right)^{2}-2PT_{h}\alpha\left(R+\gamma\right)\left(R+\tau\right)}}.\label{eq:DelMaxEfiNEHL}
\end{equation}

By solving the equations system formed by Eqs. \ref{eq:AlfMaxEfiNEHL} and \ref{eq:DelMaxEfiNEHL}, we obtain the values of $\alpha$=$\alpha_{R}^{M \eta}$ and $\delta$= $\delta_{R}^{M \eta}$ that would represent the operation mode of WT plant under the maximum efficiency regime. Thus, there exists a particular value of $\alpha^{*}=\alpha(R,P,\eta)$ for any $\alpha$ bounded between $\alpha_{ZII,ZIII}$ (see Eq. \ref{eq:CAlp23NEHL}). The operation mode reported in Almaraz II plant will always be stayed in $ZII$. However, if $\alpha>\alpha_{\eta R}$ then it will be located in $ZIII$. Analogously, there are conditions for $\alpha$ and $\delta$, so that the reported operation mode during the running of the power plants, represent the maximum power output regime. Thus, by replacing Eq. \ref{eq:ahMaxPotNEHL} in Eq. \ref{eq:Pot}, we get a new mathematical expression for $\delta$:

\begin{equation}
\delta_{R}^{MP}=\frac{\alpha_{R}^{MP}\left[\tau+R\left(1-\eta\right)+\left(\eta-2\right)\sqrt{R\tau}\right]}{\eta\left(R+\gamma\right)\left(1-\tau\right)},\label{eq:delmaxpotNEHL}
\end{equation}
in the same way, by replacing Eq. \ref{eq:ahMaxPotNEHL} in Eq. \ref{eq:Efi}, we have,

\begin{equation}
    \alpha_{R}^{MP}=\frac{P\left(R+\gamma\right)\sqrt{R\tau}}{T_h\left(\sqrt{R\tau}-R\right)\left(\sqrt{R\tau}-\tau\right)}.\label{eq:alfmaxpotNEHL}
\end{equation}

The lack of information, in the reported data during the operation of power plants, restricts the number of energetic favorable configurations, i.e, there are plants that operate in energetically unprofitable zones ($ZI$ and $ZIII$). To identify the operation zone of each analyzed plant, we use the generalization parameter of the ecological function (Eq. \ref{eq:DefEcoGen}) as the parametric variable that allows to establish a relation between the height reduced temperature, power output and efficiency. By using Eqs. \ref{eq:PotNEHL} and \ref{eq:DisNEHL} in the expression for ecological function (Eq. \ref{eq:DefEcoGen}) and by taking the derivative with respect to the high reduced temperature, we get:

\begin{equation}
    a_{h}^{ME_G}=\frac{\gamma(1+\epsilon\tau)+\sqrt{R\tau(1+\epsilon)(1+\epsilon\tau)}}{(R+\gamma)(1+\epsilon\tau)}.\label{eq:ahMaxEcoGNEHL}
\end{equation}
This optimal high reduced temperature is related to an $\epsilon$-parameter, which maximize the compromise function, as we can see in Fig. \ref{fig:G0new} a) where the maxima are in different optimal operation zones.

In Eq. \ref{eq:ahMaxEcoGNEHL}, we have $a_h=a_h(\epsilon)$; that is, $\epsilon$ characterizes the operation zones analogously to the high reduced temperature. By replacing Eq. \ref{eq:ahMaxEcoGNEHL}, into lower Ineq. \ref{eq:LahNEFC} and solving with Eq. \ref{eq:ahMaxPotNEHL}, we get the bounds in terms of $\epsilon$,

\begin{equation}
    \frac{\tau-R}{R-\tau^2}<\epsilon<0,\label{eq:limepsZ1NEHL}
\end{equation}
this inequality guarantees the operation mode is located in $ZI$. In the analogous way, by using Eqs. \ref{eq:ahMaxPotNEHL} and \ref{eq:ahMaxEfiNEHL}, we get
\begin{equation}
    0\leq\epsilon\leq\frac{\rho_{\eta R}^{2}-2R\alpha\tau \rho_{\eta R}+R\tau\left[\tau\left(R\alpha^{2}+2\alpha\kappa\delta+2\kappa^{2}\delta^{2}\right)-\tau^{2}\left(\alpha+\kappa\delta\right)^{2}-\kappa^{2}\delta^{2}\right]}{\tau\left\{ R\kappa^{2}\delta^{2}-R^{2}\rho_{\eta R}+2R\alpha\tau \rho_{\eta R}+R\tau\left(\tau\left[\alpha^{2}\left(1-R\right)+2\alpha\kappa\delta+\kappa^{2}\delta^{2}\right]-2\delta\kappa\left[\alpha+\delta\kappa\right]\right)\right\} }.\label{eq:LimEpsZIINEHL}
\end{equation}

This inequality guarantees that the operation mode is now in $ZII$. Finally, by solving with Eq. \ref{eq:ahMaxEfiNEHL} and replacing Eq. \ref{eq:ahMaxEcoGNEHL} into upper bound of Ineq. \ref{eq:LahNEFC}, we get

\begin{equation}
    \frac{\rho_{\eta R}^{2}-2R\alpha\tau \rho_{\eta R}+R\tau\left[\tau\left(R\alpha^{2}+2\alpha\kappa\delta+2\kappa^{2}\delta^{2}\right)-\tau^{2}\left(\alpha+\kappa\delta\right)^{2}-\kappa^{2}\delta^{2}\right]}{\tau\left\{ R\kappa^{2}\delta^{2}-R^{2}\rho_{\eta R}+2R\alpha\tau \rho_{\eta R}+R\tau\left(\tau\left[\alpha^{2}\left(1-R\right)+2\alpha\kappa\delta+\kappa^{2}\delta^{2}\right]-2\delta\kappa\left[\alpha+\delta\kappa\right]\right)\right\} }<\epsilon<\frac{R-\tau}{\tau\left(1-R\right)},\label{eq:LimEpsZIIINEHL}
\end{equation}
as in the previous Ineqs. (\ref{eq:limepsZ1NEHL} and \ref{eq:LimEpsZIINEHL}) this expression also guarantees that the operation mode will be located inside of $ZIII$ with $\kappa=R+\gamma$.

On the other hand, when the characteristic curve between power vs efficiency is analized, we get for a given power there are two values of the $\epsilon$ parameter. When these values are used in Eqs. \ref{eq:PotNEHL} and \ref{eq:EfiNEHL}, we get particular values for the  processes variables $P$ and $\eta$. By replacing Eq. \ref{eq:ahMaxEcoGNEHL} in the expression for power output (see Eq. \ref{eq:PotNEHL}) and by solving for $\epsilon$, we obtain:

\begin{equation}
\epsilon_{PR(1,2)}=\frac{\left(1+\tau\right)n_{\epsilon_P1}\pm\left(1-\tau\right)\left[P\left(R+\gamma\right)-T_h\alpha\left(R+\tau\right)\right]\sqrt{n_{\epsilon_P1})}}{2\tau\left[T_h^2\alpha^2\left(R-\tau^2\right)\left(1-R\right)+2PT_h\alpha\left(R+\gamma\right)\left(R+\tau\right)-P^2\left(R+\gamma\right)^2\right]},\label{eq:EPsNEHL}
\end{equation}
where $\epsilon_{PR1}$ is associated to $+$ sign and $\epsilon_{PR2}$ to $-$ sign, besides
\begin{equation}
    n_{\epsilon_P1}=P^2\left(R+\gamma\right)^2+T_h^2\alpha^2\left(R-\tau\right)^2-2PT_h\alpha\left(R+\gamma\right)\left(R+\tau\right).
\end{equation}

In analogous way, after replacing Eq. \ref{eq:ahMaxEcoGNEHL} into  Eq. \ref{eq:EfiNEHL} for efficiency and solving for $\epsilon$, we obtain

\begin{equation}
   \epsilon_{\eta R(1,2)}=\frac{R^{2}\left(1+\tau\right)\left[\alpha\left(1-\eta\right)+\delta\eta\left(1-\tau\right)^{2}\right]+n_{\epsilon\eta1}\mp\left(1-\tau\right)\left\{ R\left[\alpha\left(1-\eta\right)-\delta\eta\left(1-\tau\right)\right]-\gamma\delta\eta\left(1-\tau\right)+\alpha\tau\right\} \rho_{\epsilon\eta1}}{2\tau\left\{ \alpha^{2}\tau\left[\tau+\left(1-\eta\right)^{2}\right]-\gamma\delta^{2}\eta^{2}\left(1-\tau\right)^{2}\left(1+\tau\right)-\alpha\delta\eta\left[\gamma\left(1-\eta\right)-\tau\right]\left(1-\tau\right)^{2}\right\}},\label{EEsNEHL}
\end{equation}
with $\epsilon_{\eta R1}$ is associated to $-$ sign and $\epsilon_{\eta R2}$ to $+$ sign. Where $n_{\epsilon\eta1}$ and $\rho_{\epsilon\eta1}$ are given by:

\begin{equation}
    n_{\epsilon\eta1}=2R\left\{ \gamma\delta^{2}\left(1-\tau\right)^{2}\left(1+\tau\right)-\alpha^{2}\tau\left[\tau+\left(1-\eta\right)^{2}\right]+\alpha\delta\eta\left[\gamma\left(1-\eta\right)-\tau\right]\left(1-\tau\right)^{2}\right\} +\left(1+\tau\right)\left[\alpha\tau-\gamma\delta\eta\left(1-\tau\right)\right]^{2}
\end{equation}
and
\begin{equation}
    \rho_{\epsilon\eta1}=\sqrt{\left[\gamma\delta\eta-R\alpha+R\eta\left(\alpha+\delta\right)\right]^{2}-2\left\{ R\alpha^{2}-\alpha\eta\left[R\alpha-\delta\left(1-R\right)\left(R+\gamma\right)\right]+\delta\eta^{2}\left(R+\gamma\right)\left[\gamma\delta+R\left(\alpha+\delta\right)\right]\right\}}
\end{equation}

\begin{figure}
	\centering
		\includegraphics[scale=.5]{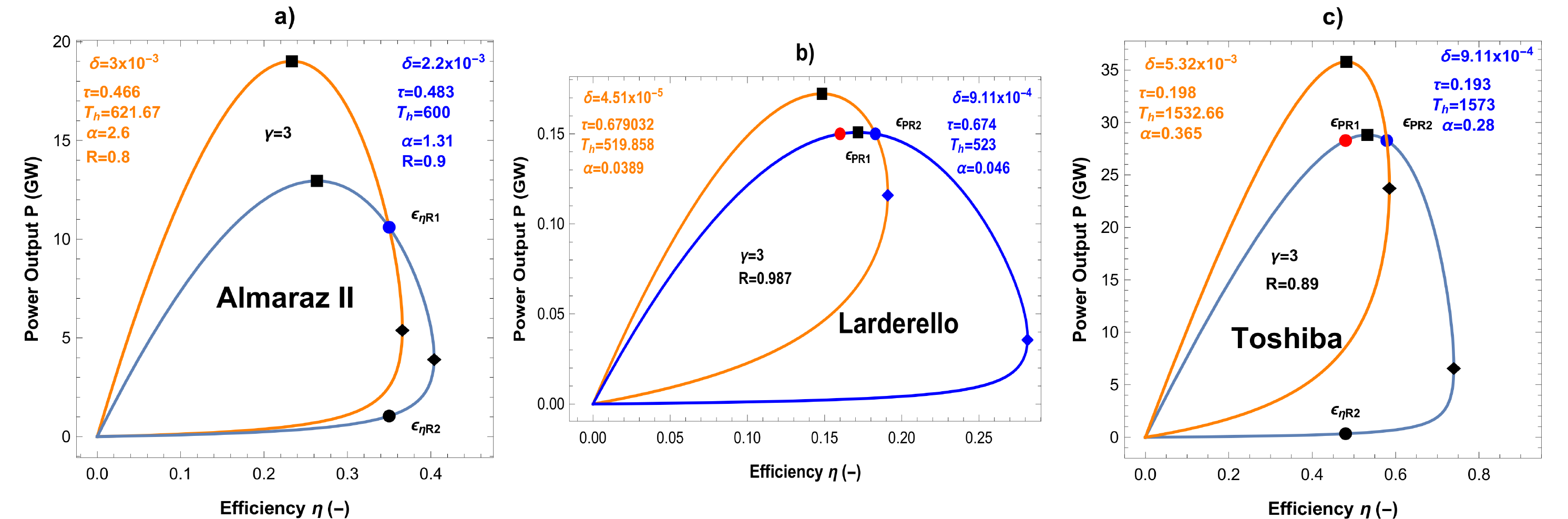}
	\caption{Curves that exemplify a configuration (blue) which contains the operation mode reported by each plant, every orange loops give an account of a specific configuration obtained from the improvement conditions under CA-$R$ case, this condition allows the plants to operate at $ZII$. In a) it is shown that Almaraz II plant can go from operating in $ZIII$ $(\epsilon_{\eta R2}>\epsilon_{MPO})$ to a point $\epsilon_{\eta R1}$ that represents its performance in $ZII$ maintaining the same efficiency and improving its power output. In b), Larderello plant works in $ZI$ ($\epsilon_{PR1}<\epsilon_{MPO}$) and the point $\epsilon_{PR2}$ symbolizes its operation in $ZII$, which has the same power output and improve its efficiency. In c) Toshiba's plant operation is schematized in $ZIII$ ($\epsilon_{\eta R2}<\epsilon_{MPO}$) and the point $\epsilon_{PR2}$ shows its performance in $ZII$ improving both the power as efficiency.}
	\label{fig:G05}
\end{figure}

Then, the operating modes that lie within a configuration curve can be specified via the $\epsilon$-parameter (see Fig. \ref{fig:G05}). And according to each operation zone, they must be specified with some of the values for $\epsilon$ (see Eqs. \ref{eq:EPsNEHL} and \ref{EEsNEHL}). 

In order to obtain the new configuration of those power plants that are not initially working in $ZII$, for example, Almaraz II and Larderello (see Table 2), we also use the following relationship between the high reduce temperatures and the external ones \cite{Levario19}, to ensures the new point lies on both the original loop and the improved one
\begin{equation}
   \frac{a_h}{a^{*}_h} = \frac{T^{*}_h}{T_h},\label{reltemp}
\end{equation}
where $T_{hw}=T_{hw}^*$ is assumed. In Table 2, it is shown to the new values for both power output and efficiency are located in the optimal region ($ZII$). They strongly depend on the restructured parameter values: $\tau^{\dagger}$, $\alpha^{\dagger}$, $\gamma^{\dagger}$, $\delta^{\dagger}$ and $R^{\dagger}$; that is, all of these new parameters define a new configuration (loop) for each power plant.

\begin{table}
\begin{centering}
{\scriptsize{}}%
\begin{tabular}{|c|c|c|c|c|c|c|c|c|c|c|c|}
\hline 
{\scriptsize{}P.P} & \multicolumn{4}{c|}{{\scriptsize{}Reported Operation Mode }} & \multicolumn{2}{c|}{{\scriptsize{}Operation Mode in ZII}} & {\scriptsize{}$T_{h}^{*}/T_{h}$} & \multicolumn{4}{c|}{{\scriptsize{}Restructuring Operation Mode}}\tabularnewline
\hline 
\multirow{4}{*}{{\scriptsize{}A}} & {\scriptsize{}$\alpha[GW/K]$} & {\scriptsize{}$R$} & {\scriptsize{}$P[GW]$} & {\scriptsize{}$\epsilon_{\eta R2}\notin ZII$} & {\scriptsize{}$\epsilon_{\eta R1}\in ZII$} & {\scriptsize{}$P[GW]$} & \multirow{4}{*}{{\scriptsize{}0.964}} & \multirow{2}{*}{{\scriptsize{}$T_{h}^{*} [K]$}} & {\scriptsize{}$\alpha^{\dagger}[GW/K]$} & {\scriptsize{}$R^{\dagger}$} & {\scriptsize{}$\epsilon^{\dagger}\in ZII$}\tabularnewline
 & {\scriptsize{}1.31} & {\scriptsize{}0.9} & {\scriptsize{}1.044} & {\scriptsize{}6.778} & {\scriptsize{}0.778} & {\scriptsize{}10.762} &  &  & {\scriptsize{}2.6} & {\scriptsize{}0.8} & {\scriptsize{}1.287}\tabularnewline
\cline{2-7} \cline{3-7} \cline{4-7} \cline{5-7} \cline{6-7} \cline{7-7} \cline{10-12} \cline{11-12} \cline{12-12} 
 & {\scriptsize{}$\delta[GW/K]$} & {\scriptsize{}$\tau$} & {\scriptsize{}$\eta$} & {\scriptsize{}$a_{h}\notin ZII$} & {\scriptsize{}$a_{h}^{*}\in ZII$} & {\scriptsize{}$\eta$} &  & \multirow{2}{*}{{\scriptsize{}621.67}} & {\scriptsize{}$\delta^{\dagger}[GW/K]$} & {\scriptsize{}$\tau^{\dagger}$} & {\scriptsize{}$a_{h}^{\dagger}\in ZII$}\tabularnewline
 & {\scriptsize{}$2.2\times10^{-3}$} & {\scriptsize{}0.483} & {\scriptsize{}0.35} & {\scriptsize{}0.997} & {\scriptsize{}0.961} & {\scriptsize{}0.35} &  &  & {\scriptsize{}$3\times10^{-3}$} & {\scriptsize{}0.466} & {\scriptsize{}0.982}\tabularnewline
\hline 
\multirow{4}{*}{{\scriptsize{}L}} & {\scriptsize{}$\alpha[GW/K]$} & {\scriptsize{}$R$} & {\scriptsize{}$P[GW]$} & {\scriptsize{}$\epsilon_{PR1}\notin ZII$} & {\scriptsize{}$\epsilon_{PR2}\in ZII$} & {\scriptsize{}$P[GW]$} & \multirow{4}{*}{{\scriptsize{}0.994}} & \multirow{2}{*}{{\scriptsize{}$T_{h}^{*} [K]$}} & {\scriptsize{}$\alpha^{\dagger}[GW/K]$} & {\scriptsize{}$R^{\dagger}$} & {\scriptsize{}$\epsilon^{\dagger}\in ZII$}\tabularnewline
 & {\scriptsize{}0.039} & {\scriptsize{}0.987} & {\scriptsize{}0.15} & {\scriptsize{}-0.081} & {\scriptsize{}0.093} & {\scriptsize{}0.15} &  &  & {\scriptsize{}0.039} & {\scriptsize{}0.987} & {\scriptsize{}0.764}\tabularnewline
\cline{2-7} \cline{3-7} \cline{4-7} \cline{5-7} \cline{6-7} \cline{7-7} \cline{10-12} \cline{11-12} \cline{12-12} 
 & {\scriptsize{}$\delta[GW/K]$} & {\scriptsize{}$\tau$} & {\scriptsize{}$\eta$} & {\scriptsize{}$a_{h}\notin ZII$} & {\scriptsize{}$a_{h}^{*}\in ZII$} & {\scriptsize{}$\eta$} &  & \multirow{2}{*}{{\scriptsize{}519.858}} & {\scriptsize{}$\delta^{\dagger}[GW/K]$} & {\scriptsize{}$\tau^{\dagger}$} & {\scriptsize{}$a_{h}^{\dagger}\in ZII$}\tabularnewline
 & {\scriptsize{}$4.51\times10^{-5}$} & {\scriptsize{}0.675} & {\scriptsize{}0.16} & {\scriptsize{}0.954} & {\scriptsize{}0.96} & {\scriptsize{}0.183} &  &  & {\scriptsize{}$4.51\times10^{-5}$} & {\scriptsize{}0.679} & {\scriptsize{}0.973}\tabularnewline
\hline 
\end{tabular}{\scriptsize\par}
\par\end{centering}
\caption{\label{tab:ResPlanNEHL}Comparison between the reported operation mode and the restructuring operation mode for the Almaraz II (A) and Larderello (L) power plants (P.P) for the CA-$R$ case.}
\end{table}

\subsubsection{CA-$r$ case}
\begin{figure}
	\centering
		\includegraphics[scale=.75]{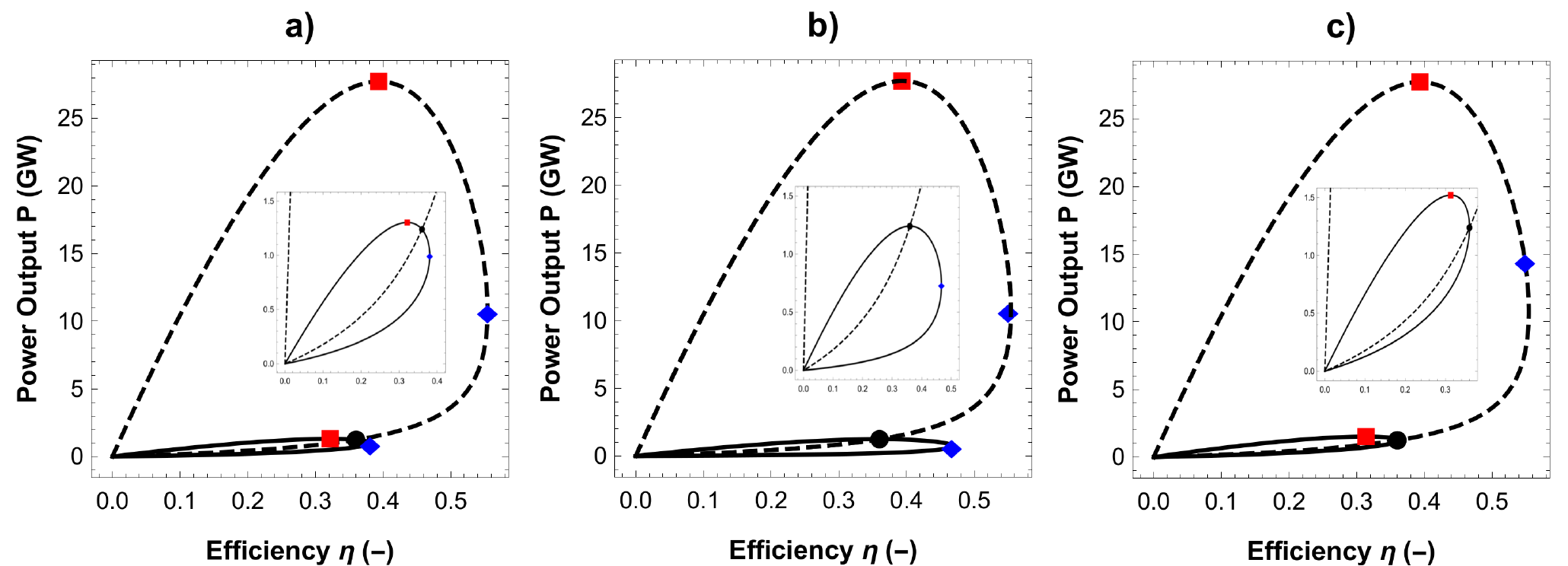}
	\caption{Loops representing different configurations for the operation of West Thurrock (WT) plant. In a) the existence of two configurations is shown, one operation mode is located in $ZII$ (solid line) and the other one belongs to $ZIII$ (dashed line). In b) there is a configuration for which the operation mode reported by WT represents the maximum power output regime under the appropriate conditions (solid line). Finally, in c) the same operation mode is observed, representing the maximum efficiency regime under other conditions (solid line).}
	\label{fig:G05a}
\end{figure}

Another way to find the restructuring conditions for the ''optimal'' operation of power plants is through the CA-$r$ case. Similarly, as in the previous section and considering the new set of variables $\alpha$, $\gamma$, $r$ and $\delta$, we have two different configurations for West Thurrock plant (Fig. \ref{fig:G05a}). In the first one (dashed line, Fig. \ref{fig:G05a}b), $\alpha=0.82 GW/K$, $\gamma=3$, $r=0.0005$ and $\delta=2.43 MW/K$ and the operation mode is in $ZIII$. While in the second configuration: $\alpha=0.0012 GW/K$, $\gamma=3$, $r=0.0046$ and $\delta=1.13 MW/K$, the operation mode is located in $ZII$. This model also shows the possibility of performing transitions between zones $ZII$ and $ZIII$. Anew, the variation of the control parameters determine the heat engine performance but, the generated loops are no equivalent to those of the CA-$R$ case. In order to the reported operation mode of WT (Table \ref{tab:DataPlant}) represent the $MP$ regime, it is necessary that  $\alpha=0.04GW/K$, $\gamma=3$ and $\delta=272.76KW/K$ (Fig.\ref{fig:G05a}b, solid line), while the $M\eta$ regime needs $\alpha=0.045GW/K$, $\gamma=3$ and  $\delta=1.96KMW/K$ (Fig. \ref{fig:G05a}c, solid line). In addition, these values belong to the intervals given by Ineqs. \ref{eq:rAIHL23} and \ref{eq:lAIHL23}. Thus, we find conditions for $\alpha$ and $\delta$ that allow themselves to reach $MP$ and $M\eta$ operation modes. The result is an equation system that arises from replacing Eq. \ref{eq:ahMaxPIHL} into Eq. \ref{eq:Pot}  in the case of $MP$ regime, the $\alpha$ value must be:

\begin{equation}
\alpha_{r}^{MP}=\frac{P\left[1+\gamma\left(1-r\right)\right]}{T_{h}\left[1-r\gamma-2\sqrt{\tau}+\tau\left(1-r\right)\right]}\label{eq:alpMaxPotIHL}
\end{equation}
and
\begin{equation}
\delta_{r}^{MP}=\frac{\alpha_{r}^{MP}\left[\left(1-r\gamma\right)\left(1-\eta\right)-\sqrt{\tau}\left(2-\eta\right)+\tau\left(1-r\right)\right]}{\eta\left(1-\tau\right)\left[1+\gamma\left(1-r\right)\right]}.\label{eq:DelMaxPotIHL}
\end{equation}

In analogous way by replacing Eq. \ref{eq:ahMaxPIHL} into \ref{eq:Efi} ($M\eta$ regime), the system of equations to solve is:
\begin{equation}
\alpha=\frac{\delta\eta\left[1+\left(1-r\right)\right]\left(1-\tau\right)\left[1-r\gamma\left(1-\eta\right)-\eta+\tau\left(1-r\right)+2\sqrt{\tau\left(1-\eta\right)}\right]}{r^{2}\left[\gamma\left(1-\eta\right)+\tau\right]^{2}+\left(1-\eta-\tau\right)^{2}-2r\left[\gamma\left(1-\eta\right)+\tau\right]\left(1-\eta+\tau\right)}\label{eq:alpMaxEfiIHL}
\end{equation}
and,
\begin{equation}
\delta=\alpha\frac{P^{2}n_{\delta1}^{2}\left[r\left(\gamma-\tau\right)+\tau-1\right]+T_{h}\alpha n_{\delta2}\left(\rho_{\eta \delta1}+T_{h}\alpha n_{\delta3}\right)-Pn_{\delta1}\left[1+\tau+r\left(\gamma+\tau\right)\right]\left(\rho_{\eta \delta1}+2T_{h}\alpha n_{\delta3}\right)}{2\left(1-\tau\right)n_{\delta1}\left\{ P^{2}n_{\delta1}^{2}-2PT_{h}\alpha n_{\delta1}\left[1+\tau-r\left(\gamma+\tau\right)\right]+T_{h}^{2}\alpha^{2}n_{\delta2}\right\} }\label{eq:DelMaxEfiIHL}
\end{equation}
where,

\begin{equation}
n_{\delta1}=1+\gamma\left(1-r\right),\label{eq:nDel0}
\end{equation}

\begin{equation}
n_{\delta2}=\left(1-\tau\right)\left(r\gamma+\tau-r\tau-1\right)^{2}-2r\left(1+\tau\right)\left(\gamma+\tau\right)+r^{2}\left(\gamma+\tau\right)^{2},\label{eq:nDel1}
\end{equation}

\begin{equation}
n_{\delta3}=r\gamma+\tau-r\tau-1\label{eq:nDel2}
\end{equation}
and 
\begin{equation}
\rho_{\eta \delta1}=\sqrt{\left\{ P\left[1+\gamma\left(1-r\right)\right]-T_{h}\alpha\left(1-r\gamma\right)\right\} ^{2}-2T_{h}\alpha\tau\left\{ T_{h}\alpha+\left[P\left(1-r\right)+r\alpha T_{h}\right]\left[1+\gamma\left(1-r\right)\right]\right\} -T_{h}^{2}\alpha^{2}\tau^{2}\left(1-r\right)^{2}}.\label{eq:raEtarIHL}
\end{equation}
After solving them (Eqs. \ref{eq:alpMaxEfiIHL} and \ref{eq:DelMaxEfiIHL}), we obtain the values: $\alpha$=$\alpha_{r}^{M \eta}$ and $\delta$= $\delta_{r}^{M \eta}$, which allow certain power plants to achieve the maximum efficiency regime.

It is a fact not all of the power plants have the same conditions to work in the ''optimal'' operation zone ($ZII$). The optimal conditions is found again through the ecological function's generalization parameter (Eq. \ref{eq:DefEcoGen}), which establishes two functional relations ($\epsilon_{(Pr,\eta r)1}$ and $\epsilon_{(Pr,\eta r)2}$) between $a_{h}^{ME_{G}}$ value and the control parameters ($\tau$, $\gamma$ and $r$).  Now, by using Eqs. \ref{eq:PotIHL} and \ref{eq:DisIHL} in the expression for ecological function (Eq. \ref{eq:DefEcoGen}) and by taking the derivative with respect to the high reduced temperature, we get:

\begin{equation}
a_{h}^{ME_{G}}=\frac{\gamma\left(1+\epsilon\tau\right)+\sqrt{\tau\left(1+\epsilon\right)\left(1+\epsilon\tau\right)}}{\left(1+\epsilon\tau\right)\left[1+\gamma\left(1-r\right)\right]}.\label{eq:ahMaxEcoGIHL}
\end{equation}
which can be obtained analogously to the CA-$R$ case.

Analogously to the CA-$R$ case, but now these limits for the values of $\epsilon$-parameter are obtained with the help of Eqs. \ref{eq:limahIHL}, \ref{eq:ahMaxPotNEHL} and \ref{eq:ahMaxEfiIHL}. Therefore, in this case, the conditions to identify the operation zones $ZI$, $ZII$ and $ZIII$ are given by: 
\begin{equation}
\frac{2\left\{ 2-n_{i}-\tau\left[2+n_{i}-\tau\right]+r^{2}\left[\gamma+\tau\right]^{2}+r\left[\gamma+\tau\right]\left[n_{i}-2\left(1+\tau\right)\right]\right\} }{\tau\left[3+n_{i}-\tau+r\left(\gamma+\tau\right)\right]\left[1+n_{i}-\tau+r\left(\gamma+\tau\right)\right]}<\epsilon<0,\label{eq:Eps1IHL}
\end{equation}
\begin{equation}
0\leq\epsilon\leq\epsilon_{II}\label{eq:Eps2IHL}
\end{equation}
and 
\begin{equation}
\epsilon_{II}<\epsilon<\frac{2\left\{ 1+n_{i}+\tau\left[n_{i}+\tau-2\right]+r^{2}\left[\gamma+\tau\right]^{2}-r\left[\gamma+\tau\right]\left[2+n_{i}+2\tau\right]\right\} }{\tau\left\{ 3+n_{i}+\tau-r\left[\gamma+\tau\right]\right\} \left\{ 1-n_{i}-\tau+r\left[\gamma+\tau\right]\right\} }\label{eq:Eps3IHL}
\end{equation}
where $\epsilon_{II}$ is:

\begin{equation}
    \epsilon_{II}=\frac{\alpha\left\{2\rho_{\eta r}-\delta\left[1-\tau\right]\left[1+\gamma\left(1-r\right)\right]\left[1+\tau-r\left(\gamma+\tau\right)\right]-\alpha\tau\left[1-\tau-r^{2}\left(\gamma+\tau\right)-r\left(1+\gamma+2\tau\right)\right]\right\} }{\left\{ \rho_{\eta r}+\delta\left[1-\tau\right]\left[1+\gamma\left(1-r\right)\right]-\alpha\tau\left[2-r\right]\right\} \left\{\rho_{\eta r}+\delta\left[1-\tau\right]\left[1+\gamma\left(1-r\right)\right]-r\alpha\tau\right\} }.\label{EMEIHL}
\end{equation}
\begin{figure}
	\centering
		\includegraphics[scale=.7]{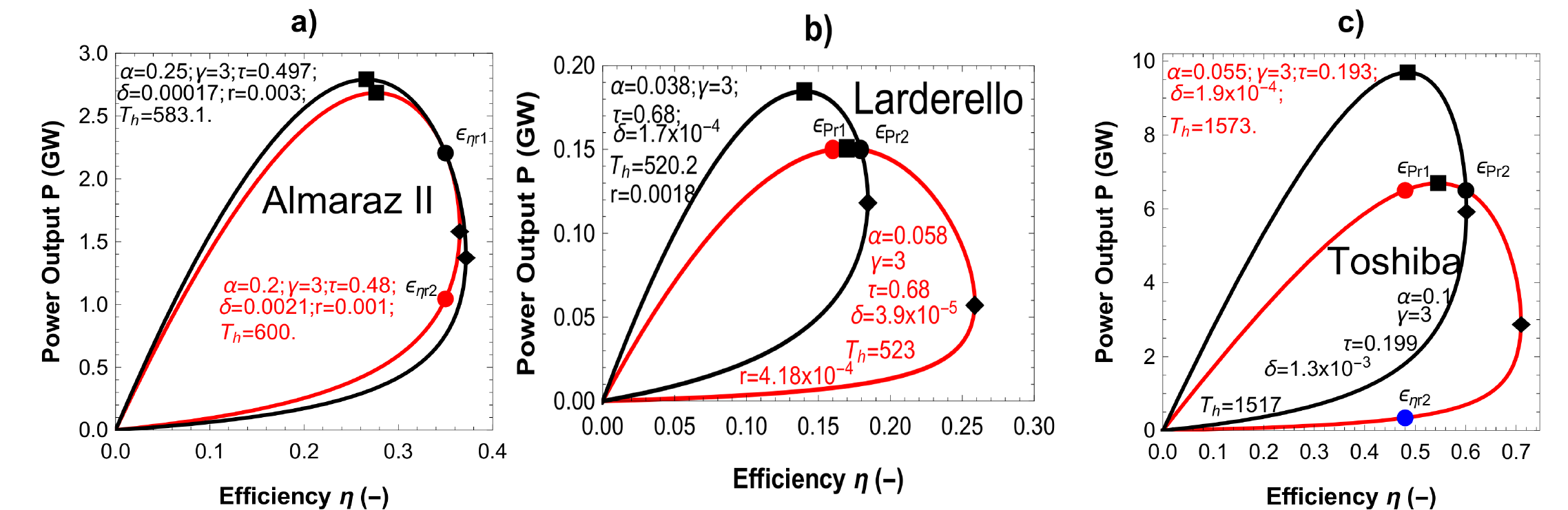}
	\caption{Curves that exemplify a configuration (red) which contains the operation mode reported by each plant, every black loops give an account of a specific configuration obtained from the improvement conditions under CA-$r$ case, this condition allows the plants to operate at $ZII$. In a) it is shown that Almaraz II can go from operating in $ZIII$ $(\epsilon_{\eta r2}>\epsilon_{MPO})$ to a point $\epsilon_{\eta r1}$ that represents its performance in $ZII$ maintaining the same efficiency and improving its power output. In b), Larderello works in $ZI$ ($\epsilon_{Pr1}<\epsilon_{MPO}$) and the point $\epsilon_{Pr2}$ symbolizes its operation in $ZII$, which has the same power output and improve its efficiency. In c) Toshiba's operation is schematized in $ZIII$ ($\epsilon_{\eta r2}<\epsilon_{MPO}$) and the point $\epsilon_{Pr2}$ shows its performance in $ZII$ improving both the power and efficiency.}
	\label{fig:G06}
\end{figure}

Likewise, to get the parametric relations $\epsilon=\epsilon(P,r)$, we replaced Eq. \ref{eq:ahMaxPIHL} into Eqs. \ref{eq:Pot} and \ref{eq:Efi}, in order to get (see Fig. \ref{fig:G06}):

\begin{equation}
\epsilon_{Pr(1,2)}=-\left(1+\tau\right)\rho_{\eta \delta1}\frac{\rho_{\eta \delta1}\mp\left(1-\tau\right) \left\{ P\left[1+\gamma\left(1-r\right)\right]+T_{h}\alpha\left[r\left(\gamma+\tau\right)-\left(1+\tau\right)\right]\right\} }{2\tau\left\{ P\left[1+\gamma\left(1-r\right)\right]+T_{h}r\alpha\left(\gamma+\tau\right)\right\} \left\{ P\left[1+\gamma\left(1-r\right)\right]+T_{h}\alpha\left[r\left(\gamma+\tau\right)-2\left(1+\tau\right)\right]\right\}}.\label{eq:EpsPIHL}
\end{equation}

Where $\epsilon_{Pr1}$ is associated to $-$ sign and $\epsilon_{Pr2}$ to $+$ sign. While replacing Eq. \ref{eq:ahMaxEfiIHL} into Eq. \ref{eq:Pot} and Eq. \ref{eq:Efi}. Now, we obtain $\epsilon=\epsilon(\eta,r)$,

\begin{equation}
\epsilon_{\eta r(1,2)}=-\frac{n_{\delta1}^{2}\delta^{2}\eta^{2}\left(1-\tau\right)^{2}\left(1+\tau\right)+n_{\epsilon1}+n_{\epsilon2}-\left( 1-\tau\right) \left\{ \delta\eta n_{\delta1}\left(1-\tau\right)\mp\alpha\left[1-\eta+\tau-r\left(\tau+\gamma\left[1-\eta\right]\right)\right]\right\} \rho_{\eta \delta2}}{2\tau\left[ \alpha\left(2-r\gamma\right)\left(1-\eta\right)-\delta\eta n_{\delta1}\left(1-\tau\right)+\alpha\tau\left(2-r\right)\right] \left\{r\gamma\left[\delta\eta-\alpha\left(1-\eta\right)\right]-\delta\eta\left(1+\gamma\right)\left(1-\tau\right)-r\tau\left(\alpha+\gamma\delta\eta\right)\right\}}\label{eq:EpsEtaIHL}
\end{equation}
where $\epsilon_{\eta r1}$ is associated to $-$ sign and $\epsilon_{\eta r2}$ to $+$ sign, besides
\begin{equation}
n_{\epsilon1}=-\alpha^{2}\left\{ \left(2r\left[\gamma\left(1-\eta\right)+\tau\right]\left(1+\tau-\eta\right)+\left[\gamma\left(1-\eta\right)+\tau\right]\right)\left(1+\tau\right)+\left(1-\tau\right)\left[\tau^{2}-\left(1-\eta\right)^{2}\right]\right\} \label{eq:nEps1IHL}
\end{equation}

\begin{equation}
n_{\epsilon2}=-2\alpha\delta\eta\left[1+\gamma\left(1-r\right)\right]\left[\tau\left(1-r\right)+\left(1-r\gamma\right)\left(1-\eta\right)\right]\left(1-\tau^{2}\right)\label{eq:nEps2IHL}
\end{equation}

\begin{equation}
\rho_{\eta \delta2}=\sqrt{n_{\delta1}^{2}\delta^{2}\eta^{2}\left(1-\tau\right)^{2}+2\alpha n_{\delta1}\delta\eta\left(1-\tau\right)\left[\left(1-\eta\right)\left(1-r\gamma\right)+\tau\left(1-r\right)\right]+n_{\epsilon3}}\label{eq:rEtarEfiIHL}
\end{equation}
and
\begin{equation}
n_{\epsilon3}=\alpha^{2}\left\{\left(1-\eta-\tau\right)^{2} -2r\left[\tau+\gamma\left(1-\eta\right)\right]\left(\tau+1-\eta\right)+r^{2}\left[\gamma\left(1-\eta\right)+\tau\right]^{2}\right\} \label{eq:nEps3IHL}
\end{equation}

The obtained expression for $\epsilon$ contains the construction elements of the energy converter ($\alpha$, $\delta$, $\gamma$, $T_{h}$ and $\tau$), as well as elements that describe a specific operation mode ($P$ and $\eta$). Analogously to Table 2, in Table 3 we show for the CA-$r$ case, the values adopted by the parameters for both the original and the restructured operation modes.

\begin{table}
\begin{centering}
{\scriptsize{}}%
\begin{tabular}{|c|c|c|c|c|c|c|c|c|c|c|c|}
\hline 
{\scriptsize{}P.P} & \multicolumn{4}{c|}{{\scriptsize{}Reported Operation Mode }} & \multicolumn{2}{c|}{{\scriptsize{}Operation Mode in ZII}} & {\scriptsize{}$T_{h}^{*}/T_{h}$} & \multicolumn{4}{c|}{{\scriptsize{}Restructuring Operation Mode}}\tabularnewline
\hline 
\multirow{4}{*}{{\scriptsize{}A}} & {\scriptsize{}$\alpha[GW/K]$} & {\scriptsize{}$r$} & {\scriptsize{}$P[GW]$} & {\scriptsize{}$\epsilon_{\eta r2}\notin ZII$} & {\scriptsize{}$\epsilon_{\eta r1}\in ZII$} & {\scriptsize{}$P[GW]$} & \multirow{4}{*}{{\scriptsize{}0.972}} & \multirow{2}{*}{{\scriptsize{}$T_{h}^{*}[K]$}} & {\scriptsize{}$\alpha^{\dagger}[GW/K]$} & {\scriptsize{}$r^{\dagger}$} & {\scriptsize{}$\epsilon^{\dagger}\in ZII$}\tabularnewline
 & {\scriptsize{}0.2} & {\scriptsize{}0.001} & {\scriptsize{}1.044} & {\scriptsize{}4.835} & {\scriptsize{}1.022} & {\scriptsize{}2.2} &  &  & {\scriptsize{}0.25} & {\scriptsize{}0.003} & {\scriptsize{}1.08}\tabularnewline
\cline{2-7} \cline{3-7} \cline{4-7} \cline{5-7} \cline{6-7} \cline{7-7} \cline{10-12} \cline{11-12} \cline{12-12} 
 & {\scriptsize{}$\delta[GW/K]$} & {\scriptsize{}$\tau$} & {\scriptsize{}$\eta$} & {\scriptsize{}$a_{h}\notin ZII$} & {\scriptsize{}$a_{h}^{*}\in ZII$} & {\scriptsize{}$\eta$} &  & \multirow{2}{*}{{\scriptsize{}538.1}} & {\scriptsize{}$\delta^{\dagger}[GW/K]$} & {\scriptsize{}$\tau^{\dagger}$} & {\scriptsize{}$a_{h}^{\dagger}\in ZII$}\tabularnewline
 & {\scriptsize{}$2.1\times10^{-3}$} & {\scriptsize{}0.48} & {\scriptsize{}0.35} & {\scriptsize{}0.981} & {\scriptsize{}0.953} & {\scriptsize{}0.35} &  &  & {\scriptsize{}$1.7\times10^{-4}$} & {\scriptsize{}0.497} & {\scriptsize{}0.957}\tabularnewline
\hline 
\multirow{4}{*}{{\scriptsize{}L}} & {\scriptsize{}$\alpha[GW/K]$} & {\scriptsize{}$r$} & {\scriptsize{}$P[GW]$} & {\scriptsize{}$\epsilon_{Pr2}\notin ZII$} & {\scriptsize{}$\epsilon_{Pr1}\in ZII$} & {\scriptsize{}$P[GW]$} & \multirow{4}{*}{{\scriptsize{}0.769}} & \multirow{2}{*}{{\scriptsize{}$T_{h}^{*} [K]$}} & {\scriptsize{}$\alpha^{\dagger}[GW/K]$} & {\scriptsize{}$r^{\dagger}$} & {\scriptsize{}$\epsilon^{\dagger}\in ZII$}\tabularnewline
 & {\scriptsize{}0.058} & {\scriptsize{}$4.18\times10^{-4}$} & {\scriptsize{}1.15} & {\scriptsize{}-0.072} & {\scriptsize{}0.081} & {\scriptsize{}1.15} &  &  & {\scriptsize{}0.038} & {\scriptsize{}0.0018} & {\scriptsize{}0.721}\tabularnewline
\cline{2-7} \cline{3-7} \cline{4-7} \cline{5-7} \cline{6-7} \cline{7-7} \cline{10-12} \cline{11-12} \cline{12-12} 
 & {\scriptsize{}$\delta[GW/K]$} & {\scriptsize{}$\tau$} & {\scriptsize{}$\eta$} & {\scriptsize{}$a_{h}\notin ZII$} & {\scriptsize{}$a_{h}^{*}\in ZII$} & {\scriptsize{}$\eta$} &  & \multirow{2}{*}{{\scriptsize{}520.2}} & {\scriptsize{}$\delta^{\dagger}[GW/K]$} & {\scriptsize{}$\tau^{\dagger}$} & {\scriptsize{}$a_{h}^{\dagger}\in ZII$}\tabularnewline
 & {\scriptsize{}$3.9\times10^{-5}$} & {\scriptsize{}0.68} & {\scriptsize{}0.16} & {\scriptsize{}0.75} & {\scriptsize{}0.904} & {\scriptsize{}0.24} &  &  & {\scriptsize{}$1.7\times10^{-4}$} & {\scriptsize{}0.68} & {\scriptsize{}0.973}\tabularnewline
\hline 
\end{tabular}{\scriptsize\par}
\par\end{centering}
\caption{\label{tab:ResPlanIHL}Comparison between the reported operation mode and the restructuring operation mode for the Almaraz II (A) and Larderello (L) power plants (P.P) for the CA-$r$ case.}
\end{table}

\section{Conclusions}
As is well known, the characteristic loops of performance of a real thermal engines are not produced in the analysis of the power output versus efficiency curves of an endoreversible engine model. To emulate one of these loops, it is necessary to incorporate within the CA--$R$ case, not only the irreversibility parameter $R$ but also a heat leak term between the two temperature external reservoirs. However, in the CA--$r$ case these loops arise without the need to incorporate this heat leakage, due to the exchange of energy and its surroundings. Therefore, a more complete model must incorporate such a dissipative element. In this work, we characterize the restrictions of the different parameters involved in the irreversible cases that permit the physical configuration achievable for the energy converter. Likewise, we have shown that a particular mode of operation corresponds to a set of configurations that can be found from two different ways. The parameters $\alpha$ and $\delta$ generate a greater effect on the behavior of the converter, due to their relationship with heat exchangers that allow the heat flows with the surroundings. In addition, the constraints on $R$ and $r$ parameters were analyzed. These parameters can roughly estimate a certain irreversibility degree of each energy converter here studied. Besides, we obtain the relationship between the parameters associated with the design and construction  of the energy converter model and their internal irreversibility degrees. On the other hand, we show the importance of the $\epsilon$-parameter to improve conditions in the performance of some power plants in both CA-$R$ and CA-$r$ cases. Besides, by using the $\epsilon$-parameter in both maximum power output and maximum efficiency regimes allow us to classify the operation mode along the characteristic loops in the configuration space. In addition, through $\epsilon$-parameter we can obtain the achievable conditions in order to an energy converter can operate in the optimal performance region; that is, at $ZII$ (see Figs. \ref{fig:G05} and \ref{fig:G06}). Finally, as can be observed in Figs. \ref{fig:G05} and \ref{fig:G06}, as well as in Tables 2 and 3, the generalization parameter $\epsilon$ allows us to find improvement conditions so that power plants operate with better energy performance.

\section*{Acknowledgement}
The authors want to thanks to Professor F. Angulo-Brown for his recommendations to improve this manuscript. And also thanks to the anonymous reviewer for his relevant comments that made it possible to improve the article.

\end{document}